\documentclass[sigconf, 10pt]{acmart}
\usepackage{graphicx}
\usepackage{float}
\usepackage{subfigure}
\usepackage{subcaption}
\usepackage{pifont}
\usepackage{setspace}
\usepackage{enumitem}
\usepackage{placeins}
\usepackage{amsthm}
\usepackage[linesnumbered,ruled,vlined]{algorithm2e}
\usepackage{multirow}

\copyrightyear{2025}
\acmYear{2025}
\setcopyright{acmlicensed}\acmConference[ACM MOBICOM '25]{The 31st Annual International Conference on Mobile Computing and Networking}{November 4--8, 2025}{Hong Kong, China}
\acmBooktitle{The 31st Annual International Conference on Mobile Computing and Networking (ACM MOBICOM '25), November 4--8, 2025, Hong Kong, China}
\acmDOI{10.1145/3680207.3723493}
\acmISBN{979-8-4007-1129-9/2025/11}

\begin{document}

\title{D$^{2}$MoE: Dual Routing and Dynamic Scheduling for Efficient On-Device MoE-based LLM Serving}
\renewcommand{\shorttitle}{D$^2$MoE}

\author{Haodong Wang}
\email{hwanghb@connect.ust.hk}
\orcid{0009-0001-8977-850X}
\affiliation{
  \institution{Hong Kong University of Science and Technology}
  \city{Hong Kong}
  \country{China}
}

\author{Qihua Zhou}
\authornote{Corresponding author.}
\email{qihuazhou@szu.edu.cn}
\affiliation{
  \institution{College of Computer Science and Software Engineering, Shenzhen University}
  \city{Shenzhen}
  \country{China}}

\author{Zicong Hong}
\authornotemark[1] 

\email{congcong@ust.hk}
\affiliation{
  \institution{Hong Kong University of Science and Technology}
  \city{Hong Kong}
  \country{China}
}

\author{Song Guo}
\email{songguo@cse.ust.hk}
\affiliation{
  \institution{Hong Kong University of Science and Technology}
  \city{Hong Kong}
  \country{China}}

\renewcommand{\shortauthors}{Haodong Wang et al.}

\begin{abstract}
The mixture of experts (MoE) model is a sparse variant of large language models (LLMs), designed to hold a better balance between intelligent capability and computational overhead. Despite its benefits, MoE is still too expensive to deploy on resource-constrained edge devices, especially with the demands of on-device inference services. 
Recent research efforts often apply model compression techniques, such as quantization, pruning and merging, to restrict MoE complexity. 
Unfortunately, due to their predefined static model optimization strategies, they cannot always achieve the desired quality-overhead trade-off when handling multiple requests, finally degrading the on-device quality of service.
These limitations motivate us to propose the D$^2$MoE, an algorithm-system co-design framework that matches diverse task requirements by dynamically allocating the most proper bit-width to each expert. 
Specifically, inspired by the nested structure of matryoshka dolls, we propose the \textit{matryoshka weight quantization} (MWQ) to progressively compress expert weights in a bit-nested manner and reduce the required runtime memory. On top of it, we further optimize the I/O-computation pipeline and design a heuristic scheduling algorithm following our \textit{hottest-expert-bit-first} (HEBF) principle, which maximizes the expert parallelism between I/O and computation queue under constrained memory budgets, thus significantly reducing the idle temporal bubbles waiting for the experts to load.
Evaluations on real edge devices show that D$^2$MoE improves the overall inference throughput by up to 1.39$\times$ and reduces the peak memory footprint by up to 53$\%$ over the latest on-device inference frameworks, while still preserving comparable serving accuracy as its INT8 counterparts.  

\end{abstract}



\maketitle

\begin{figure}[t]
  \centering
  \includegraphics[width=\linewidth]{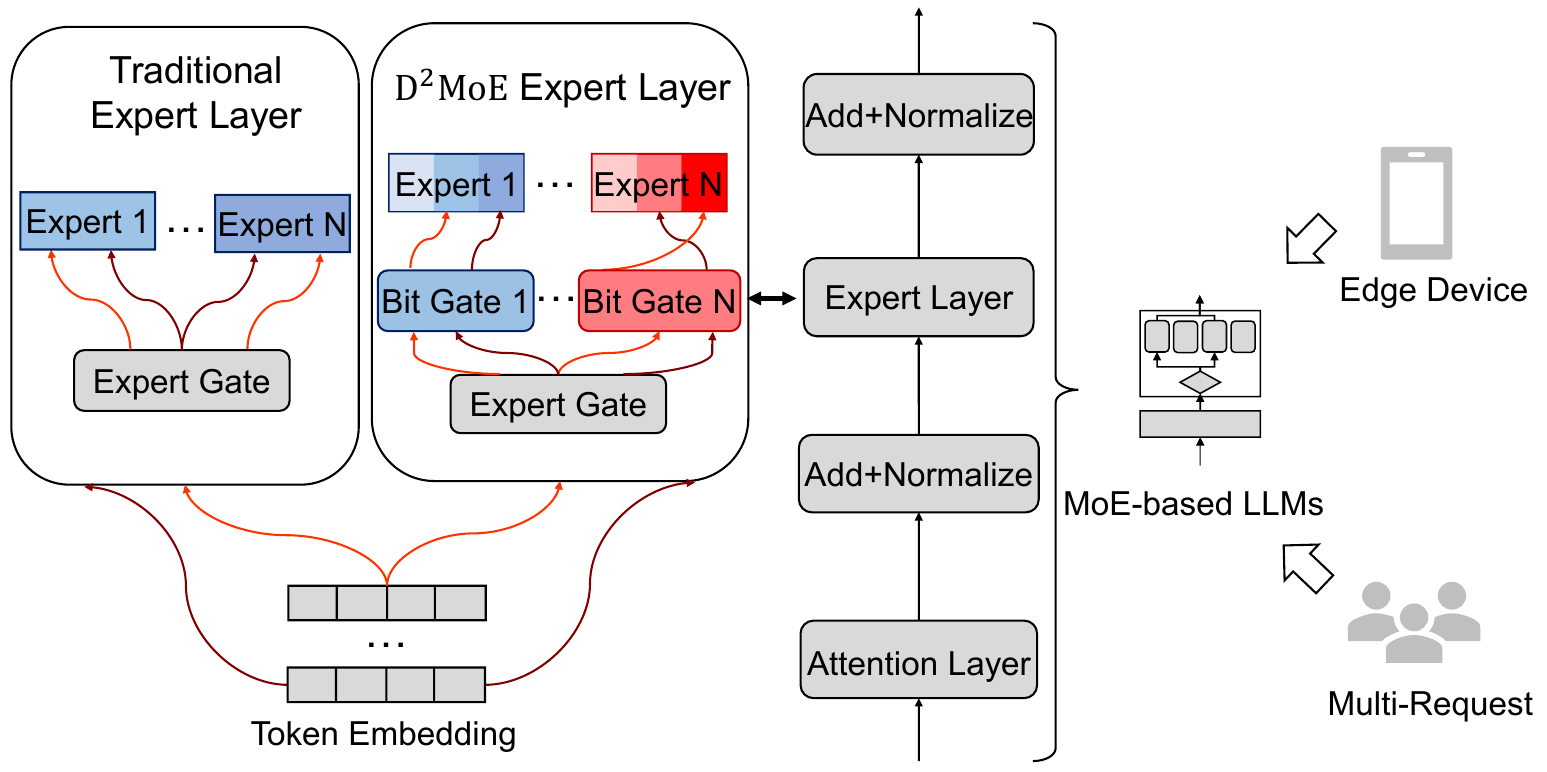}
  \caption{Traditional MoE single routing (expert ID only) vs. our D$^2$MoE dual routing (ID and bit-wdith).}\label{D2MoE_overview}
\end{figure}

\section{INTRODUCTION}

The Large Language Models (LLMs)~\cite{llama2, LLM_1, LLM_opt, LLM_mixtural} are increasingly embedded in our work and everyday activities, supporting tasks like summarization, code completion, and decision making~\cite{codeChen2021, LLM_task_1, LLM_task_copilot, LLM_task_chatgpt}. To further scale these models without incurring prohibitive training and inference costs, recent research has introduced Mixture-of-Experts (MoE) models, where feedforward network (FFN) layers are often replaced with MoE layers. These MoE layers enable sparse activation for each token, dynamically routing computation to a subset of model components (i.e., experts). As model sizes continue to grow and task complexity increases, the sparse activation architecture of MoE models becomes increasingly critical. However, compared to other LLMs, MoE-based models face more severe memory bottlenecks during inference due to their large parameter sizes. This challenge is particularly pronounced in resource-constrained edge environments. For instance, when stored in a compact float16 format, the parameters of Mixtral 8$\times$7B~\cite{LLM_mixtural} require over 90GB of memory, whereas consumer-grade GPUs such as the NVIDIA RTX 3060 laptop offer only 6GB of memory, highlighting the resource limitations of edge devices.

Quantization is one of the most prominent solutions for memory optimization during LLM inference on edge devices. Applying quantization to MoE models is particularly effective, as experts account for almost 90\% of the parameters (e.g., in Mixtral-8$\times$7B, experts contribute 89.9\% of the parameters, while attention mechanisms account for less than 11.1\%). Consequently, quantizing experts significantly reduce memory usage and the size of parameter transfers between storage hierarchies during inference. Recent efforts have proposed quantization strategies tailored to MoE-based models, where bit-width is assigned to different experts based on their importance on calibration dataset and fix bit-width assignments throughout inference. For instance, EdgeMoE~\cite{Yi2023EdgeMoEFO} assigns bit-width offline by quantifying the accuracy loss of experts on a calibration dataset, while MC-MoE~\cite{huang2024mc-moe} uses expert activation frequency and confidence to design a fixed bit-width allocation strategy during inference. 

However, due to the dynamic sparse activation characteristic of MoE-based models, the importance of an expert can vary with different input tokens. Inspired by related work on dense models~\cite{raposo2024mod, gong-etal-2024-mixture}, a straightforward approach involves introducing a lightweight, trainable adapter before each transformer block to dynamically decide whether to skip the current block's computation. This approach significantly reduces overall FLOPs and accelerates inference. Specifically, the adapter dynamically allocates more computational blocks to important tokens while skipping less significant tokens, optimizing computational efficiency.
Nevertheless, incorporating \emph{dynamic bit-width selection} for activated experts inevitably increases memory overhead. This is because the weights for different bit-width in existing quantization methods are stored independently. For instance, in LLaMA-MoE~\cite{llama-moe}, when quantized using llama.cpp~\cite{LLMmobile_llamacpp}, INT4 experts require 3.81 GB of memory, while storing INT2/3/4 experts simultaneously demands 9.62 GB. This significantly increases the memory footprint in resource-constrained edge devices.

Although parameters of inactive experts can be offloaded to lower-tier storage units (e.g., CPU or SSD), the I/O latency associated with quantized experts remains substantial (e.g., at Mixtral $\times$7B, INT4 quantization experts still account for more than 70\% of the parameters), leading to significant computational bubbles caused by waiting for I/O. For example, on an NVIDIA RTX 3060, the average computation time for a single expert in LLaMA-MoE is 3.1 ms, whereas the average data transfer time is approximately 20 ms. Existing methods attempt to overlap I/O and computation by processing multiple batches simultaneously~\cite{STI_ASPLOS2023, cao2024moelightninghighthroughputmoeinference}. However, due to the token variability inherent in gating mechanisms, the total number of activated experts and their associated bit-width can increase significantly. Furthermore, when the gating mechanism selects more than one expert per token (as in Mixtral-8×7B and LLaMA-MoE), the I/O overhead for transferring expert parameters is multiplied, causing GPU to experience more frequent idle periods while waiting for expert parameters.

To tackle these challenges, we propose a \emph{dually sparsely-gated Mixture-of-Experts} paradigm. Specifically recognizing that expert importance varies dynamically with different tokens, we first introduce an end-to-end fine-tuning strategy for the gating network, called \emph{token-adaptive bit-width selection}. This strategy enables dynamic bit-width decisions for each token, achieving a better trade-off among model accuracy and peak memory usage for MoE-based LLMs. Second, we present \emph{matryoshka weight quantization} (MWQ), which compresses expert weights into a structure where bit-width can be shared hierarchically. In this approach, higher bit-width always encapsulates lower bit-width, resembling the nested structure of matryoshka dolls.\footnote{matryoshka dolls are a set of wooden dolls of decreasing size, nested within one another.} Finally, we employ a \emph{bit-width-aware I/O-compute pipeline} that dynamically reorganizes the I/O and computation order for different bit-width. This pipeline processes multiple requests in batches, optimizing the parallel efficiency of expert I/O and computation. Together, these innovations improve the overall performance and efficiency of MoE-based LLMs, particularly in resource-constrained environments.

In this paper, based on the above paradigm, we propose \textbf{D$^2$MoE}, a novel MoE-based LLM dynamic quantization framework that can determine the bit-width of each expert online dynamically for expedited, resource-efficient, and high-quality on-device inference. 
The main contributions are summarized as follows.

\begin{itemize}[leftmargin=0.2cm, itemindent=0cm]
    \item We propose a dually sparsely-gated MoE paradigm that leverages the varying importance of experts across tokens to dynamically allocate expert bit-width. This approach aims to minimize peak memory usage and reduce the I/O overhead for experts.
    
    \item We design MWQ, a multi-step quantization method that nests high-bit-width expert weights into low-bit-width weights to reduce redundant memory usage.
    
    \item We propose a fine-grained expert I/O-compute pipeline paradigm to minimize bubbles between expert I/O and computation of different bit-width and design a hottest-expert-bit-first algorithm to heuristically formulate execution plans for this paradigm.
    
    \item We implement \textbf{D$^2$MoE} and conduct extensive experiments on various edge devices (e.g., NVIDIA RTX 3060, NVIDIA Jetson AGX Orin 64G) and MoE-based LLMs. Experimental results demonstrate that D$^2$MoE achieves up to a 1.39$\times$ throughput under different memory budgets compared to state-of-the-art quantization frameworks \cite{Yi2023EdgeMoEFO, MoQE}.
\end{itemize}

\section{BACKGROUND AND MOTIVATION}
\subsection{On-Device Memory Constraints}

While many existing optimizations have focused on dynamically reducing the computational cost~\cite{2024mixtureofdepths}, memory is the real bottleneck for on-device MoE-based LLM inference. Unlike computation, which may only slow down the inference, memory is often a hard constraint that directly determines whether it is feasible to run the model. This memory constraint manifests MoE-based LLMs in the following aspects:

\textbf{\emph{Hardware.}} The growth of memory capacity in edge devices significantly lags behind that of high-performance data centers in the cloud. For instance, the memory capacity of NVIDIA’s high-performance GPUs has increased nearly 90$\times$, from the Tesla P100 in 2016 to the B200 in 2024. In contrast, the memory capacity of smartphones has only grown 6$\times$, from the iPhone 6 in 2014 to the iPhone 15 in 2023.

\textbf{\emph{Model.}} The memory requirements of MoE-based LLMs typically scale linearly with their model capacity~\cite{switchtransformer}. Consequently, the memory demands of MoE-based LLMs have grown rapidly due to the expansion of model size and the need for higher performance. MoE-based LLMs, such as Switch Transformer and Mixtral, leverage multiple expert networks to enhance model capacity, which results in a substantial increase in memory consumption. For example, deploying one of the state-of-the-art MoE models, Mixtral-8$\times$7B requires at least 90GB of memory, nearly 10$\times$ the memory capacity of the most advanced edge devices.

To bridge the gap between the limited memory capacity of edge devices and the high memory requirements of MoE-based LLMs, it is imperative to design an inference framework that prioritizes optimizing memory.
\subsection{Observation}
The above analysis highlights the necessity of optimizing memory usage by quantizing the model expert weights in edge environments. This section investigates the potential advantages of dynamically adjusting the expert bit-width to align with hardware characteristics, a pivotal consideration in the design of D$^2$MoE.

\textbf{Observation \#1}:
\textbf{Different bit-width in MoE-based LLM quantization bring different benefits in terms of accuracy-memory-latency.}
In most cases, the bit-width of quantization weights are predefined hyper-parameters that remain fixed during model inference. However, in MoE-based LLMs, varying the bit-width of different experts can provide distinct advantages in terms of accuracy, memory usage, and inference latency. As illustrated in Table \ref{trade}, the LLaMA-MoE model \cite{llama-moe} was evaluated with various quantization bit-width on an RTX 4090 GPU with 24GB of memory using the llama.cpp \cite{LLMmobile_llamacpp} inference framework. The figure demonstrates the impact on memory footprint, latency, and model accuracy under different quantization settings. Specifically, quantizing the model to INT2 compared to INT8 reduces the memory footprint by 58\% and improves latency by 30\%, but the model accuracy drops drastically by 43\%.

\textbf{Summary:} \emph{This highlights the importance of incorporating bit-width into the expert selection process for quantized MoE-based LLMs. This consideration is crucial for optimizing model performance in terms of accuracy, memory usage and latency.}

 \begin{table}[t]
  \caption{Trade-offs among different bit-width.}
  \label{trade}
  \begin{tabular}{cccc}
    \toprule
    Bit-width  & mem. (GB) & lat. (token/s) & acc. (ppl $\downarrow$) \\
    \midrule
    2 & 3.04 & 50.47 & 20.95\\
    3 & 3.80 & 45.91 & 15.10 \\
    4 & 4.48 & 43.82 & 14.72 \\
    5 & 5.10 & 40.15 & 14.63 \\
    6 & 5.60 & 37.72 & 14.62 \\
    8 & 7.24 & 35.34 & 14.55 \\
    16 & 13.60 & 23.45 & 14.55 \\
    \bottomrule
  \end{tabular}
\end{table} 

\begin{figure}[t]
  \centering
  \includegraphics[width=\linewidth]{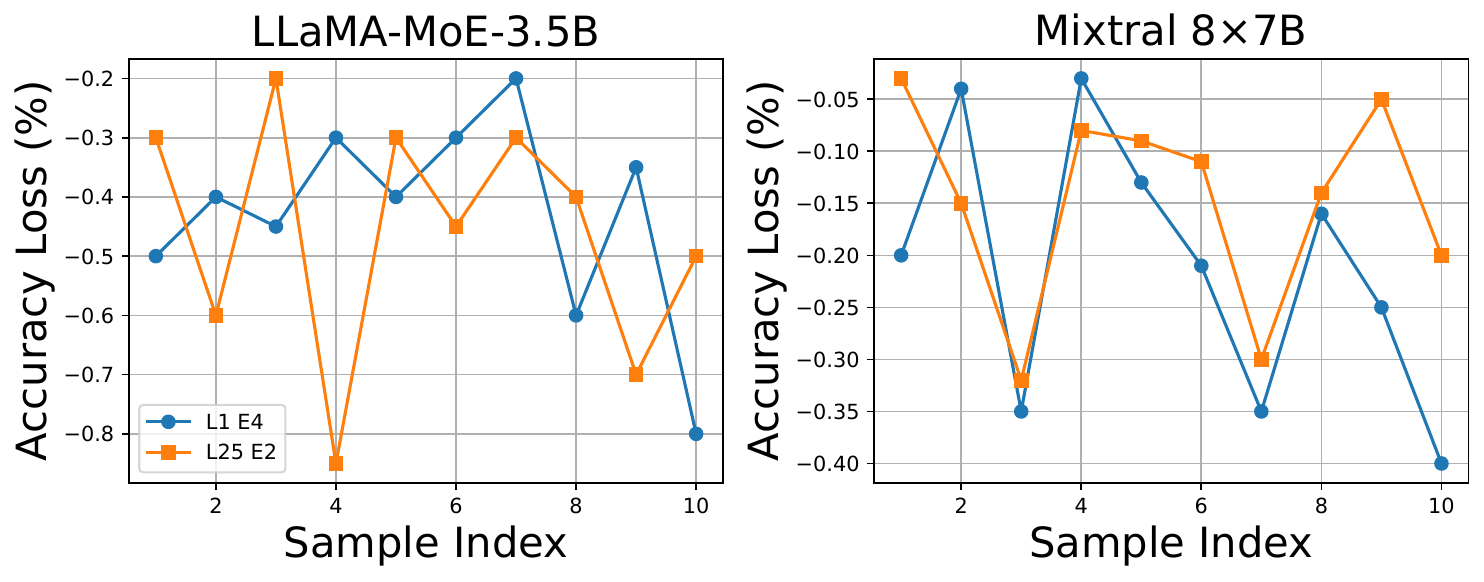}
  \caption{Accuracy loss of expert quantization to INT1 across 10 samples from the Hellaswag dataset.}\label{dynamic importance}
\end{figure}

\textbf{Observation \#2}: 
\textbf{The importance of experts changes dynamically according to different input samples.} 
Many LLM quantization methods have found that the model shards (e.g. layers and experts) show different importance to the accuracy of the model, and in order to ensure accuracy and reduce redundant I/O, they allocate higher bit-width to critical slices through offline profiling \cite{Yi2023EdgeMoEFO, STI_ASPLOS2023, mobicom_expert_quant}. However, we observe that the importance of different experts changes dynamically with the input sample. As shown in Figure \ref{dynamic importance}, we quantize expert 4 in layer 1 and expert 2 in layer 25 to INT1 while keeping other experts' bit-width unchanged. We evaluate the accuracy loss (expert importance) of LLaMA-MoE-3.5B and Mixtral 8$\times$7B across 10 samples from the Hellaswag dataset. Our results reveal significant variability in precision loss across samples and even individual tokens. For instance, quantizing the $4^{th}$ expert in the $1^{st}$ layer to 1-bit results in a 0.5\% accuracy drop on sample 1 for LLaMA-MoE-3.5B and a 0.2\% drop for Mixtral 8$\times$7B.

\textbf{Summary:} \emph{This highlights that the importance of each expert varies for different input samples, and thus, dynamically adjusting the expert bit-width to find the optimal setting for the activated experts is crucial for enhancing model performance.}

\begin{figure}[t]
  \centering
  \includegraphics[width=\linewidth]{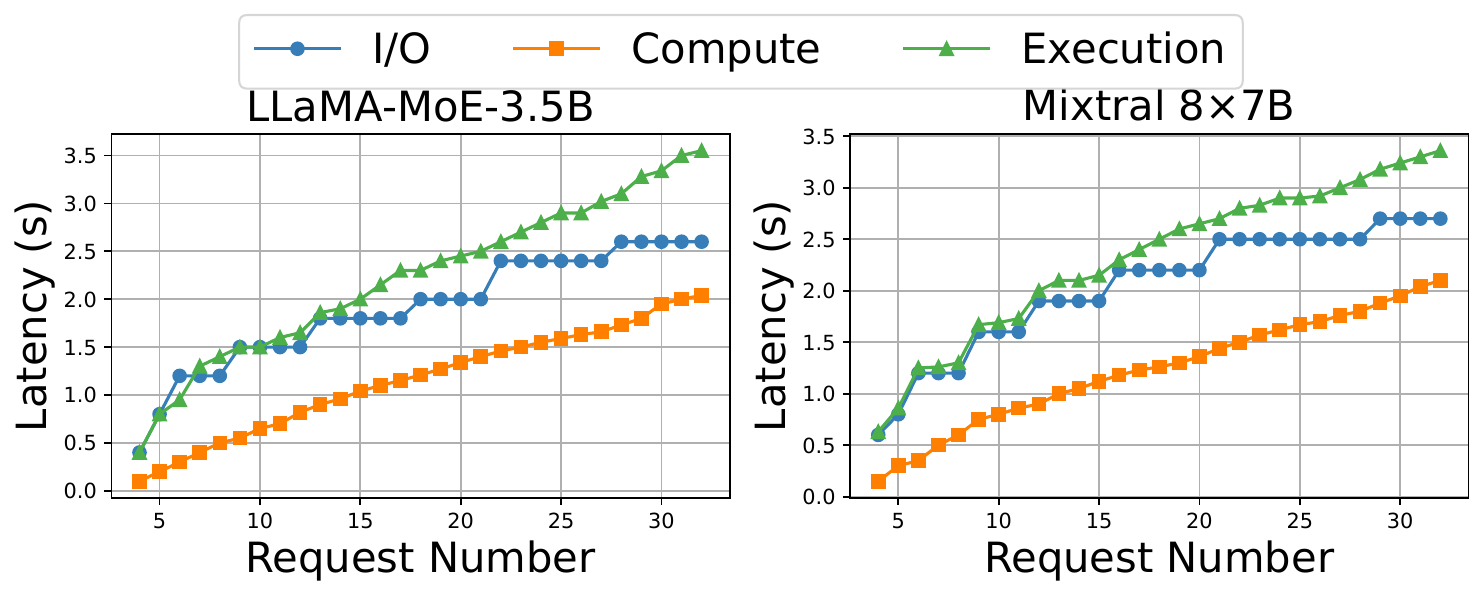}
  \caption{Comparison of expert I/O, computation, and inference latency with different request numbers.}\label{pipline bubble}
\end{figure}

\textbf{Observation \#3}: 
\textbf{Large bubble between I/O and computation of quantized experts led to substantial inference delays.}
Due to the limited memory capacity of edge devices, we adopt an on-demand method for loading activated quantized experts from disk during inference. However, as shown in Figure \ref{pipline bubble} the existing approach of loading experts in ascending order of expert IDs introduces significant bubbles between I/O and computation, leading to increased inference latency, particularly when request numbers exceeds 25. For instance, in LLaMA-MoE-3.5B with 32 requests, the expert I/O time is 2.6s, the computation time is 2.04s, and the total inference latency for the expert layer is 3.55s, which is 1.36$\times$ and 1.74$\times$ of the I/O and computation times.

\textbf{Summary:} \emph{This highlights that the I/O-compute pipeline paradigm for quantized experts create significant inefficiencies. There is a pressing need for designing scheduling plan aimed at minimizing bubbles during inference.}

\subsection{Technical Challenges}
Despite the insight that the dynamic expert bit-width routing policy is intuitive, there are still several challenges associated with implementing D$^2$MoE in complex edge environments.

\noindent\textbf{Challenge \#1. Unbalanced and inefficient bit-width selection load.} Edge devices, constrained by limited computational resources, require a trained lightweight gating network to dynamically select the appropriate bit-width for each expert without introducing significant computational overhead. The training of this gating network poses significant challenges, primarily due to two fundamental issues. Firstly, there is an imbalance in the selection of expert bit-width, as evidenced by the consistent selection of the same bit-width by numerous tokens for a specific expert \cite{switchtransformer}. Secondly, there is an irrational allocation of bit-width, where the selected expert bit-width is unable to ensure model accuracy \cite{2024mixtureofdepths}.

\noindent\textbf{Challenge \#2. High memory overhead to store multiple versions of quantized experts.}
Limited memory on edge devices is also one of the main bottleneck constrain inference performance \cite{STI_ASPLOS2023,flexnn_mobicom2024}. If the basic LLM quantization \cite{lin2023awq,xiao2023smoothquant} is used, multiple quantization models with different bit-width versions have to be deployed, further exacerbating the high memory costs associated with LLM deployment. Therefore, to make D$^2$MoE memory-efficient, a new quantization method should be devised to avoid storing multiple versions of different bit-width.

\noindent\textbf{Challenge \#3. Significant runtime overhead on weight dequantization operations.}
During model inference, the weight-only quantization approach requires the online transformation of quantized weights into the same data type as the activation for matrix computation. This dequantization operation introduces significant runtime overhead, typically accounting for 20\%-70\% of the entire inference process. To minimize the time spent on dequantization, it is crucial to design specific dequantization kernels tailored to the quantization method and aligned with the hardware characteristics.

\noindent\textbf{Challenge \#4. Lightweight and efficient online scheduling strategies address the large parallelism bubble between I/O and computation.}
Multi-requests are inherently heterogeneous and unpredictable in terms of resource and latency requirements, posing considerable challenges to LLM serving. Existing methods address these challenges by focusing on model placement policies and adaptive batch scheduling to achieve I/O-compute parallelism and reduce response latency \cite{sun2024llmserve,2024vllm}. However, quantized LLMs serving must also consider the varying bit-width selections for different requests, complicating the design of a lightweight scheduling algorithm and parallel strategy.

\section{D$^2$MoE SYSTEM DESIGN}

\subsection{System Overview}

\begin{figure}[t]
  \centering
  \includegraphics[width=\linewidth]{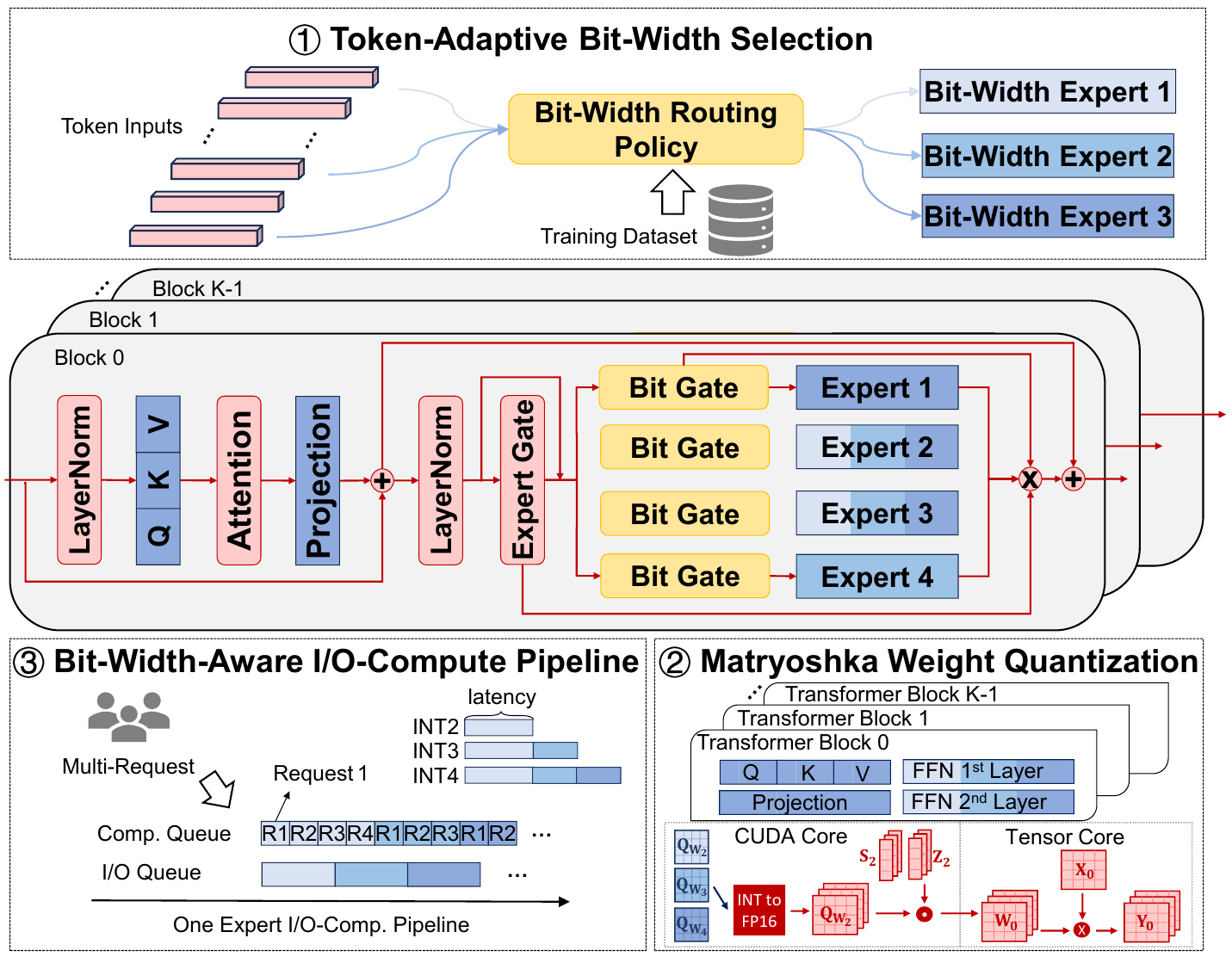}
  \caption{The architecture overview of D$^2$MoE.}\label{overview}
\end{figure}

D$^2$MoE is the first execution engine designed to enable fast inference of quantized MoE-based LLMs on edge devices. As illustrated in Figure \ref{overview}, D$^2$MoE operates through two primary stages: the offline preprocessing phase and the online execution phase.
The offline preprocessing phase, which is executed once prior to deployment, comprises two key modules: \ding{172} token-adaptive bit-width selection and \ding{173} matryoshka weight quantization. Initially, the token-adaptive bit-width selection module optimizes the bit-width allocation for different experts. Specifically, a lightweight plug-in network is trained for each expert using a generic dataset (e.g. C4 dataset) to dynamically select the bit-width for each token to achieve accuracy-memory-latency resources optimization.  Following this, the D$^2$MoE profiler applies the MWQ module to the MoE-based LLM using a small calibration dataset, effectively reducing the model's memory footprint. 

In the online execution phase, the fine-tuned, quantized MoE-based LLMs from the offline preprocessing phase are deployed onto physical edge devices. The D$^2$MoE engine then implements \ding{174} the bit-width-aware I/O-compute pipeline to manage and schedule the execution of various requests in real-time which effectively minimizes the significant idle periods between I/O and computation, thereby enhancing overall efficiency.

\subsection{Token-Adaptive Bit-Width Selection}
\begin{figure}[t]
  \centering
  \includegraphics[width=\linewidth]{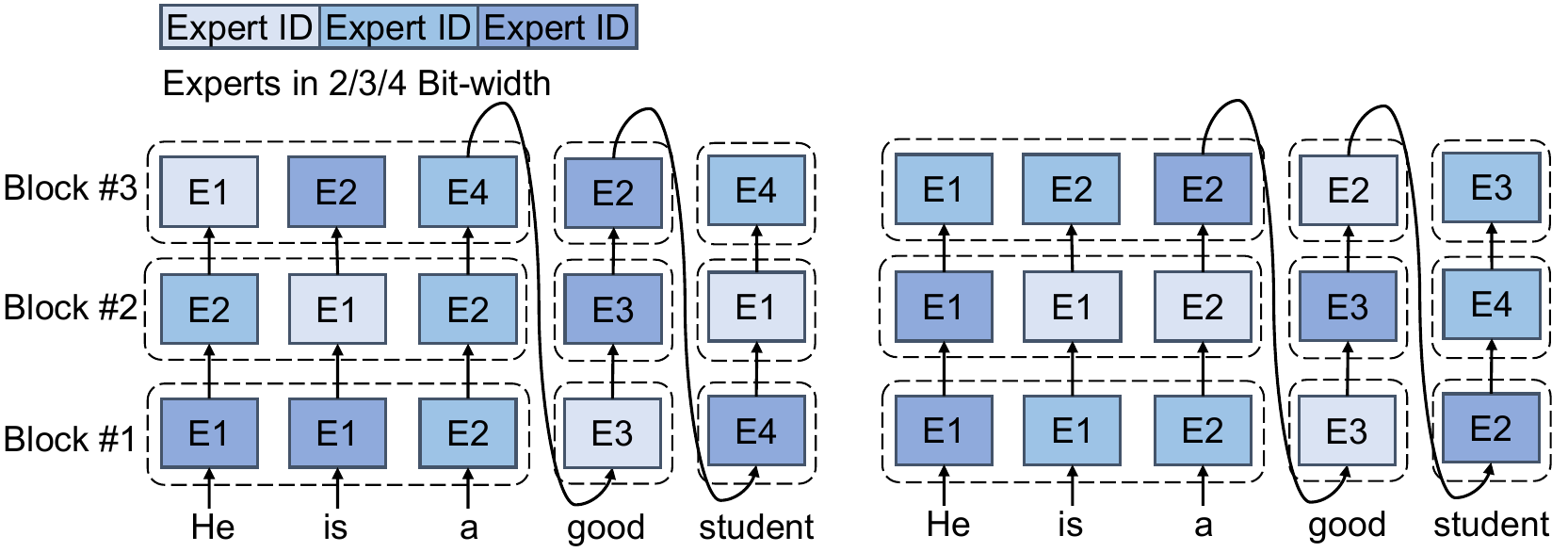}
  \caption{Comparison between fixed and dynamic bit-width allocation. }\label{token-adaptive}
\end{figure}

Inspired by MoD \cite{raposo2024mod}, the network can identify tokens critical to accuracy and assign higher bit-width activation experts accordingly. Moreover, dynamic bit-width selection aims to minimize peak memory footprint by making real-time decisions during inference. Our approach further reduces expert memory consumption while maintaining accuracy. As shown in Figure \ref{token-adaptive}, traditional methods (left) quantize experts to a hybrid (e.g., INT2/3/4) bit-width offline and keep it static during inference. In contrast, the proposed token-adaptive bit-width selection (right) achieves the same output quality with significantly lower memory usage through dynamic bit-width allocation, enabling more efficient inference.

It involves 2 steps: (1) \textit{quantized expert capacity} that balances the selection frequency of each bit-width during fine-tuning, and (2) \textit{dynamic bit-width selection loss} that optimizes the router to dynamically allocate bit-width based on the quantized expert capacity.

\textbf{Quantized expert capacity} constrains the token capacity of each expert during fine-tuning to prevent overfitting to specific token sequences. Specifically, given that the total number of tokens processed by each transformer block is $T$, we define the quantized expert capacity as $\{c_k\}_{k=1}^K$, where $\sum_{k=1}^K c_k = 1$. This formulation indicates that the maximum number of tokens assigned to the $k$-th bit-width expert during each forward propagation in fine-tuning is $c_k\cdot T$. Any tokens exceeding this capacity will skip the computation of the corresponding expert. For example, if the total number of tokens is 60 and the capacity for a particular bit-width is 0.2, then this expert can process at most 12 tokens during fine-tuning. If 14 tokens are assigned to this expert in a forward pass, 2 tokens will be randomly dropped skipping computation for these tokens at this layer.

The values of $\{c_k\}_{k=1}^K$ are predefined based on hardware constraints to optimize memory and computational efficiency and remain fixed during fine-tuning. In addition, a balanced allocation across bit-width is crucial, as higher bit-width increase memory consumption, while lower bit-width may compromise accuracy.

\begin{figure}[t]
  \centering
  \includegraphics[width=0.9\linewidth]{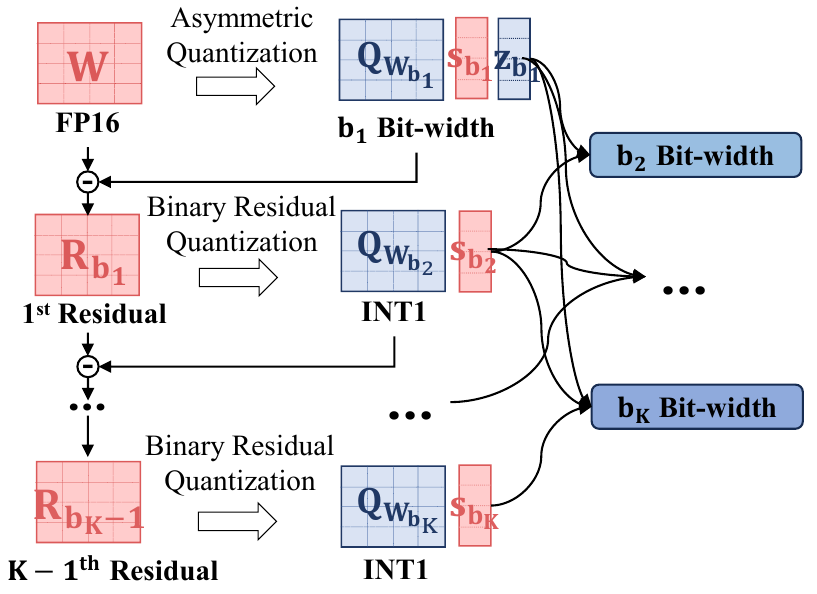}
  \caption{The workflow of MWQ.}\label{Matryoshka}
\end{figure}

\textbf{Dynamic bit-width selection loss} is introduced to fine-tune the bit-width router for selecting the optimal bit-width for activated experts. A lightweight, trainable bit-width router is placed before each expert to dynamically allocate bit-width, ensuring that higher bit-width are used for computing the most critical tokens by appropriately adjusting the logits. However, unlike the expert router, the bit-width router primarily focuses on model accuracy, which may lead to consistently favoring high bit-width due to their typically lower accuracy loss. To address this, we propose a novel bit-width balancing loss that complements the model accuracy loss to balance the selection frequency of different bit-width.  Specifically, given a list of candidate bit-width experts ($\{b_k\}^K_{k=1}$) and a batch $\mathcal{S}$ containing $T$ tokens in a forward propagation, the total loss function can be described as: 
\begin{equation}\label{1}
 Loss = \frac{1}{T}\sum_{x\in\mathcal{S}}\left(CE(p(x), q(x))+\frac{\alpha}{L}\sum_{l=1}^L\sum_{k=1}^K p_k^l(x) b_k\right).
\end{equation}
where $p_k^l(x)$ represents the probability fraction assigned by the bit-width router to the $k$-th bit-width expert at layer $l$, $p(x)$ and $q(x)$ denotes the logits of the D$^2$MoE model and the original precision model (e.g., FP16) for token $x$ after the LM head layer, respectively.

In this loss function, the first term is the cross-entropy loss, which encourages the bit-width router to prioritize higher bit-width. The second term serves as a regularization term, promoting the selection of lower bit-width to achieve a balance. Therefore, the bit-width router dynamically selects different bit-width during inference while maintaining overall model accuracy.

\subsection{Matryoshka Weight Quantization}

Token-adaptive bit-width selection effectively reduces memory overhead during inference. However, traditional quantization methods typically require storing multiple quantized versions at different bit-width, resulting in significant storage overhead. To address this challenge, inspired by the nested structure of Russian matryoshka dolls, we propose a novel multi-step quantization technique, MWQ which restructures expert weights into a nested hierarchy, embedding low bit-width weights within high bit-width weights. In the following sections, we first introduce the MWQ quantization algorithm, followed by the design of its corresponding dequantization kernel tailored to this technique.

\subsubsection{Quantization Method}\label{section 3.3.1}
Figure \ref{Matryoshka} illustrates the multi-step process of the proposed MWQ technique leveraging a list of candidate bit-width ($\{b_{k}\}_{k=1}^K$). The process begins by quantizing the weights to the minimum supported bit-width (e.g., INT2 or INT4, denoted as $b_1$) using \textit{asymmetric quantization}. Subsequently, the bit-width is iteratively increased by quantizing the residual weights through \textit{binary residual quantization} until the final $b_K$ bit-width weight is obtained. In each step, transitioning from $b_k$ to $b_{k+1}$, one additional bit-width is added along with the related scale factors.

\begin{figure}[t]
  \centering
  \includegraphics[width=\linewidth]{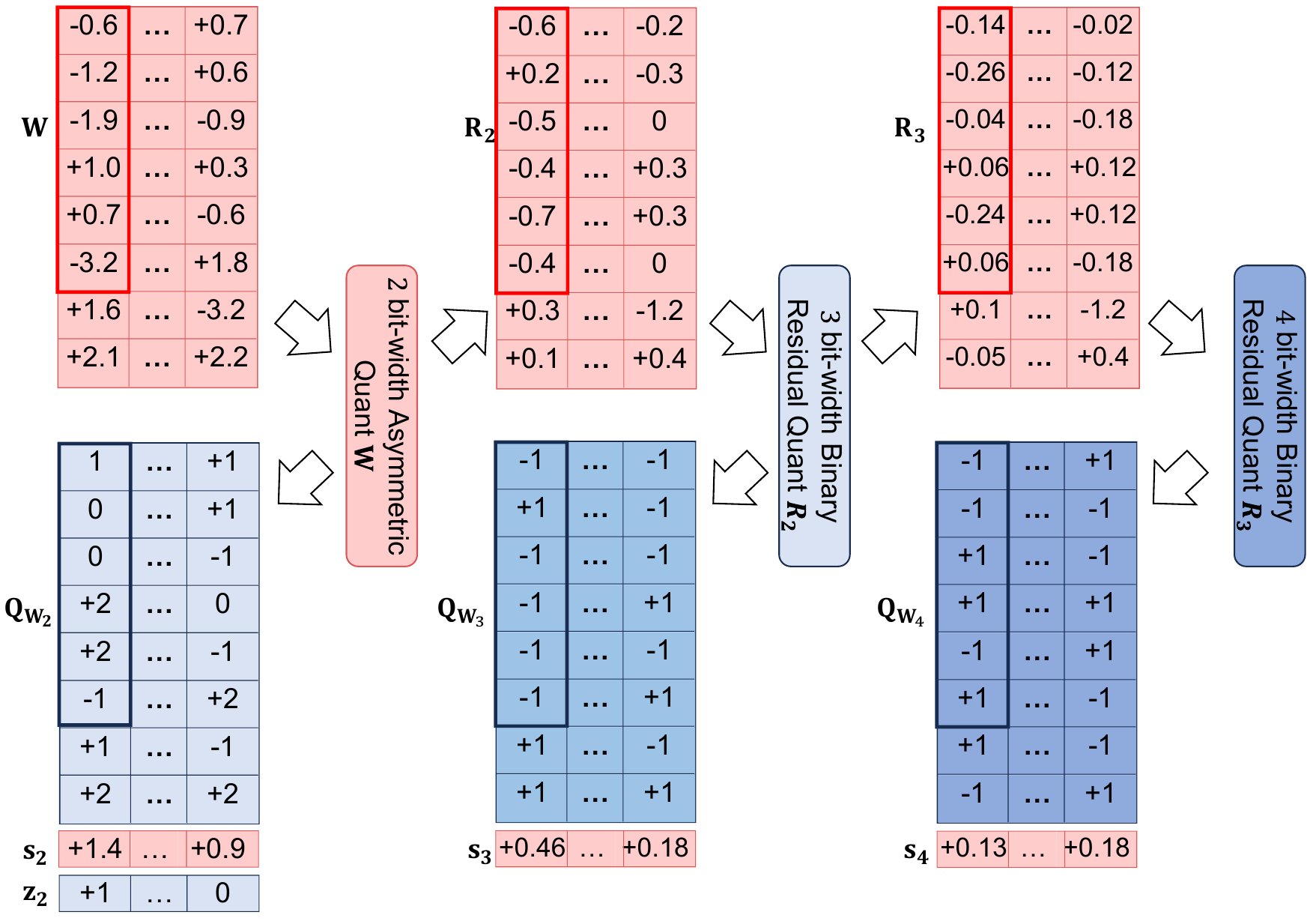}
  \caption{An example of MWQ for the weight matrix $W$ with $b_1 = 2$ and $K = 3$.}\label{Matryoshka_example}
\end{figure}

\textbf{Asymmetric Quantization.} Sparse expert weights have been shown to be robust to asymmetric quantization, particularly under low bit-width settings \cite{MoQE}. During inference, tensors are dequantized to FP16 to enable matrix multiplication with activations. To mitigate precision loss, we first apply per-group asymmetric \(b_1\) bit-width quantization. Quantization and dequantization are computed as follows:
\begin{equation}
\textbf{Q}_{\textbf{W}_{b_1}} = round(\textbf{W}/ \textbf{s}_{b_1} + \textbf{z}_{b_1}), 
\hat{\textbf{W}}_{b_1} = (\textbf{Q}_{\textbf{W}_{b_1}} - \textbf{z}_{b_1}) \cdot \textbf{s}_{b_1},
\end{equation}
where $\textbf{W} \in \mathbb{R}^{s\times h}$ represents the floating-point weight tensor, $\textbf{Q}_{\textbf{W}_{b_1}} \in \mathbb{R}^{s\times h}$ is the quantized weight tensor, and $\textbf{z}_{b_1}, \textbf{s}_{b_1} \in \mathbb{R}^{s\times h / g}$ are the zero points and scale factors for group-wise quantization. These are optimized as follows:
\begin{equation}\label{equ1}
\mathop{\arg\min\limits_{\textbf{z}_{b_1}, \textbf{s}_{b_1}}} \Vert \textbf{WX} - \hat{\textbf{W}}_{b_1} \textbf{X} \Vert_2^2,
\end{equation}
where $\textbf{X} \in \mathbb{R}^{h \times r}$ denotes the floating-point activation tensor, and $g$ is the group size. To compute weights for higher bit-width, we quantize the residuals $\textbf{R}_{b_1} = \textbf{W} - \hat{\textbf{W}}_{b_1}$.

\textbf{Binary Residual Quantization.} To ensure that low bit-width weights are subsets of higher bit-width weights while maintaining accuracy, we progressively apply per-group quantization with the binary residual approximation based on the $b_1$ bit-width quantized residuals. The binary residual quantization and dequantization are computed as:
\begin{equation}
\textbf{Q}_{\textbf{W}_{b_k}} = round(\textbf{R}_{b_{k-1}}/\textbf{s}_{b_k}), 
\hat{\textbf{Q}}_{\textbf{W}_{b_k}} = \textbf{s}_{b_k} \cdot \textbf{Q}_{\textbf{W}_{b_k}},
\end{equation}
where $k = 2, \cdots, K$, $\textbf{Q}_{\textbf{W}_{b_k}} \in \{+1, -1\}^{s \times h}$ represents the accumulated one-bit weights from $b_{k-1}$ to $b_k$, and $\textbf{s}_{b_k}$ is the per-group scale factor optimized as:
\begin{equation}\label{equ4}
\mathop{\arg\min\limits_{\textbf{s}_{b_k}}} \Vert \textbf{W} \textbf{X} - \hat{\textbf{W}}_{b_k} \textbf{X} \Vert_2^2,
\end{equation}
where $\hat{\textbf{W}}_{b_k} = \hat{\textbf{W}}_{b_1} + \sum_{i=b_2}^{b_k} \textbf{s}_{b_i} \textbf{Q}_{\textbf{W}_{b_k}}$ is the floating-point approximation of $b_k$ bit-width quantized weights. By iteratively adding low bit-width quantized weights, arbitrary bit-width quantized weights can be constructed.

\begin{algorithm}[t]
\caption{Main algorithm of MWQ}
\label{mwq}
\SetNlSty{bfseries}{}{:}   
\DontPrintSemicolon
\KwIn{Weight tensor $\textbf{W}\in\mathbb{R}^{s\times h}$, input tensor $\textbf{X}\in\mathbb{R}^{h\times r}$,  block size $\gamma$, 
 Hessian regularizer $\lambda$.} 
$\textbf{H}^c :=Cholesky ((2\textbf{X}\textbf{X}^{\mathrm{T}}+\lambda\textbf{I})^{-1})$ \;
$\textbf{Q}_{\textbf{W}_{b_i}}:= \textbf{0}_{s\times h}, i = 1, \cdots, K$\;
\For{$i < K$}{
  \For{$b = 0, \gamma, 2\gamma, \cdots$}{
        $\textbf{W}^b := \textbf{W}_{:, b: b+\gamma}$\;
    \If{i = 1}{
        $\textbf{Q}_{\textbf{W}_{b_i} :, b:b+\gamma}:=$ asym\_quant$(\textbf{W}^b)$ \;
        $\textbf{R}_{b_i}^b := \textbf{W}^b -  \hat{\textbf{W}}_{b_i}^b$ \;}
    \Else{$\textbf{Q}_{\textbf{W}_{b_i} :, b:b+\gamma}:=$ res\_quant$(\textbf{R}_i^b)$\;
    $\textbf{R}_{b_i}^b := \textbf{R}_{b_{i-1}}^b -  \hat{\textbf{W}}_{b_i}^b$\;}
    $\textbf{E}:= (\textbf{W}_{:, b: b+\gamma} - \textbf{Q}_{\textbf{W}_{b_i} :, b:b+\gamma}) / \textbf{H}^c_{b:b+\gamma, b+\gamma}$ \;
    $\textbf{W}_{:,b+\gamma:} := \textbf{W}_{:,b+\gamma:}-\textbf{E}\cdot\textbf{H}^c_{b:b+\gamma, b+\gamma:}$ \;
  }
}
\KwOut{$\{\textbf{Q}_{\textbf{W}_{b_i}}\}_{i=1}^K$}
\end{algorithm}

For example, as shown in Figure \ref{Matryoshka_example}, we first apply asymmetric quantization (group size = 6) to obtain INT2 weight $\textbf{Q}_{\textbf{W}_2}$. Next, the residual $\textbf{R}_2$ is quantized via binary residual quantization to obtain an additional 1-bit weight, forming the INT3 weight by combining $\textbf{Q}_{\textbf{W}_2}$ and $\textbf{Q}_{\textbf{W}_3}$. This process is repeated to generate another quantized weight $\textbf{Q}_{\textbf{W}_4}$, resulting in INT4 weight as the sum of all previous quantized weights.

Inspired by GPTQ \cite{gptq2022}, we further enhance the efficiency of post-training quantization by retaining only block-level compensation while eliminating column-level error corrections, ensuring the effectiveness of the MWQ strategy. Algorithm \ref{mwq} provides a detailed outline of the complete MWQ process.

\subsubsection{Dequantization Kernel}\label{section 3.3.2}
In per-group quantization, balancing accuracy enhancement with dequantization overhead is critical, yet prior studies have not facilitated efficient GEMM parallelism on GPU for dynamic bit-width. The primary performance constraint in executing dequantization for MWQ on edge device is the limited parallelism between tensor loading from various storage levels in the GPU and computations within the CUDA cores and Tensor cores. To address this, we have developed a parallel loading dequantization kernel that optimizes all levels of GPU storage.

\begin{figure}[t]
  \centering
  \includegraphics[width=\linewidth]{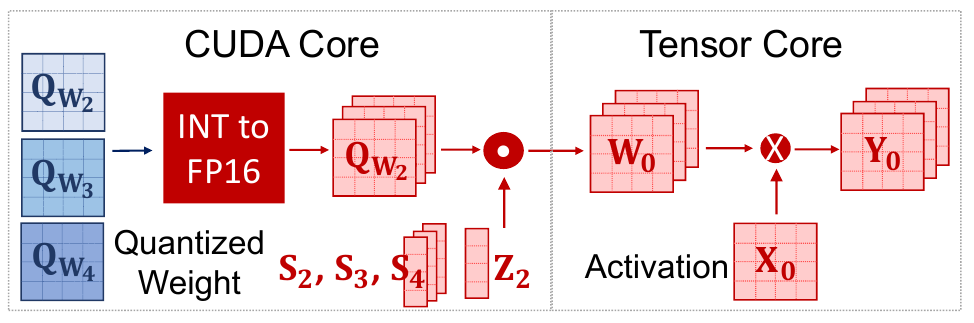}
  \caption{The dequantization overview of D$^2$MoE.}\label{de-quantization}
\end{figure}

This approach leverages a key innovation: fully overlapping tensor loading with tensor computations to simultaneously maximize bandwidth usage and computation throughput. Our method achieves loading parallelism by dynamically transferring quantized data from disk directly to GPU’s global memory, concurrently with activations moving from global memory to L2 cache. For computation parallelism, as illustrated in Figure \ref{de-quantization}, expert dequantization in the CUDA cores is synchronized with expert computation in the Tensor core. Notably, traditional bit-transpose methods from various integer formats to FP16 are inefficient; we instead employ an optimized binary operation from the Any-Precision LLM \cite{park2024anyprecision}, significantly enhancing processing speed.

\subsection{Bit-Width-Aware I/O-Compute Pipeline}
\begin{figure}[t]
\centering
\subfigure[Sequential parallelism without MWQ]{
            \centering
		\includegraphics[width=\linewidth]{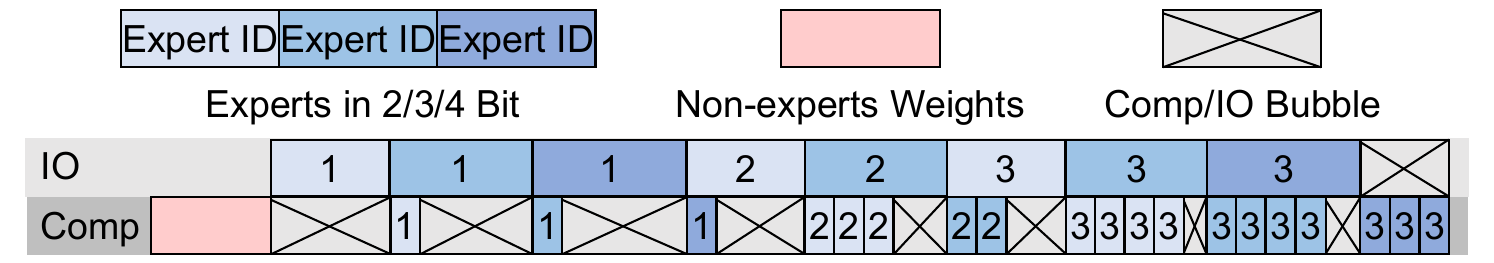}}\\
\subfigure[Sequential parallelism with MWQ]{
            \centering
		\includegraphics[width=\linewidth]{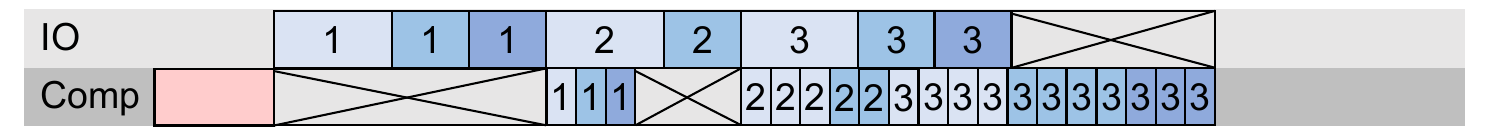}}\\
\subfigure[Fine-grained sequential parallelism with MWQ]{
            \centering
		\includegraphics[width=\linewidth]{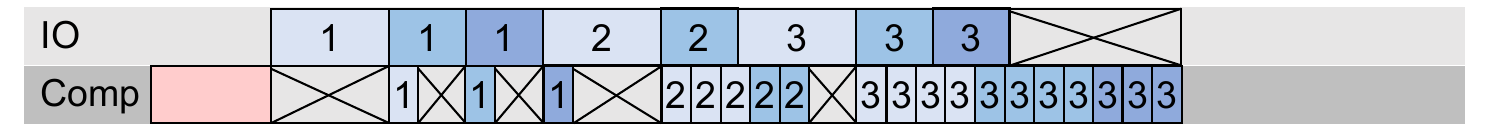}}\\
\subfigure[The optimal pipeline schedule]{
            \centering
		\includegraphics[width=\linewidth]{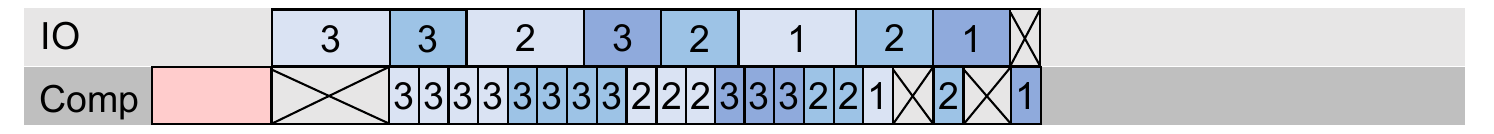}}
\caption{Comparison between different I/O-compute parallel strategies.}
\label{pipline}
\end{figure}

The nested structure of MWQ quantization reveals a limitation in the existing I/O-compute pipeline paradigm, which fails to account for scenarios where experts with different bit-width are invoked across multiple requests, leading to significant parallel bubbles. 
To be specific, Figure \ref{pipline} compares four distinct scheduling paradigm. The traditional I/O-compute execution paradigm, which does not employ MWQ, sequences the I/O and compute queue of the expert module in ascending order by expert IDs and bit-width (Figure \ref{pipline}a). 
MWQ reduces the I/O size of experts by nesting low bit-width weights within high bit-width weights, thereby increasing the utilization frequency of low bit-width weights. For instance, if three requests select Expert 2, with one selecting INT2 and two selecting INT3, MWQ ensures that all three requests call the INT2 weight (light blue, Expert 2), while the two requests requiring INT3 further call the medium blue weight (Expert 2). This nesting improves parallel efficiency by reusing low bit-width weights. However, due to sequential execution, significant parallel bubbles still occur (Figure \ref{pipline}b).
Furthermore, It has been demonstrated that MWQ is capable of performing expert I/O and computation scheduling at a fine-grained bit-width level. (Figure \ref{pipline}c). 
Ultimately, the optimal schedule (Figure \ref{pipline}d) minimize parallel bubbles by determining the execution order of experts at a fine-grained bit-width level during inference.
Therefore, $D^2$MoE employs the bit-width-aware I/O-compute pipeline paradigm to reorder the activated experts with different bit-width, thereby reducing I/O wait time. In the following, the memory budget scheduler is introduced with the aim of reducing the frequency of expert I/O. Secondly, the bit-width-aware pipeline problem will be formulated, and then the Hottest-Expert-Bit-First (HEBF) algorithm will be introduced as a solution to the pipeline problem.

\subsubsection{Memory Budget}\label{memory_budget}
To support MoE-based LLM inference in edge environments with dynamic memory constraints and reduce frequently loading experts, we introduce a memory budget $M$ during expert I/O-compute pipeline. This parameter defines the upper limit of GPU memory allocated to experts and is configurable based on the available memory resources of edge hardware. Increasing the parameter $M$ enables low bit-width weights, which are activated with greater frequency, to remain in GPU memory. Therefore, the necessity for frequent reloading is reduced.

As shown in Algorithm \ref{alg:schedule}, at each layer, we first check whether the memory required by the current expert exceeds the available budget $M$ (line 3). If the memory is sufficient, the pipeline of bit-width-aware I/O and computation is executed directly (line 9), followed by an update of the memory budget (line 10). If the memory is insufficient, high bit-width expert weights are released to free memory (lines 4-6). If the budget remains inadequate, low bit-width weights are also released as needed (lines 7-8). Finally, the pipeline is executed, and the memory budget is updated accordingly (lines 9-10).

\begin{algorithm}[t]
\caption{Memory-Budget Scheduler}
\label{alg:schedule}
\SetNlSty{bfseries}{}{:}   
\DontPrintSemicolon
\KwIn{Generate length $n$, number of layers $L$, number of bit-width $K$, available memory budget $M$.} 

\For{$i < n$}{
  \For{$j < L$}{
    \If{layers[$j$] $>$ $M$}{
        \For{$k = 0$ \KwTo $K-1$}{
            \textbf{Free}(layer[j-1][k])\;
            \textbf{Update} $M$\;
        }
        \If{layers[$j$] $>$ $M$}{
            \textbf{Free}(layer[j-1][1])}}
    \textbf{Load} and \textbf{Store} (layer[j])\;
    \textbf{Update} $M$\;
  }
}
\end{algorithm}
 
\subsubsection{Offline Profiling}
D$^2$MoE measures the following hardware capabilities of the edge device at installation time. 
\begin{itemize}[leftmargin=0.2cm, itemindent=0cm]
    \item $T_{io}(b_k)$: D$^2$MoE measures the average disk access delay for loading one expert in $b_k$ bit-width, where $b_k\in\{b_k\}_{k=1}^K$. It only has to measure one expert per bit-width because all others have the same amount of parameters.
    \item  $T_{comp}(b_k)$: D$^2$MoE calculates the average computation delay by measuring the dequantization delay for an expert with bit-width $b_k$ and the execution delay for processing a token. As the sizes of expert weights are deterministic, measuring a single expert per bit-width suffices.
\end{itemize}

The delays can be recorded offline and subsequently replayed at runtime because they are data-independent \cite{IOcomp2022} and consistently determined by the bit-width.

\subsubsection{Parallelism Planning}: In order to minimize the inference latency while satisfying the constraints, the goal of parallelism planning is to find an optimal I/O-Compute execution queue.

\textbf{Variables.} For each transformer block $l$, the set $\Omega_l$ represents the execution queue, encompassing all the experts' bit-width indices selected. The matrix $B_{j,k} \in \mathbb{R}^{N \times K}$ indicates the number of times the $k$-th bit-width of the $j$-th expert was selected, where $N$ and $K$ respectively denote the total number of experts and bit-width. For every $s\in\Omega_l$ $L(s,j,k)$ and $C(s,j,k)$ specify the start time of the $s$-th quantized expert in the execution queues and the index of this quantized expert is the $k$th bit-width of the $j$th expert. The terms $T_{io}(b_k)$ and $T_{comp}(b_k)$, previously defined, apply here as well.

\textbf{Objective.} Given that inference latency is influenced by bubbles during the parallel execution of tasks in the I/O-compute queues, we define our latency target as the difference between the total time overhead required to complete the compute queue and the load queue:
\begin{subequations}\label{eq:ctr_shale}
\begin{align*}
       \min \sum_{j=1}^{N}\sum_{k=1}^K\left(B_{j,k}T_{comp}(k)-\sigma(B_{j,k}>0) T_{io}(k)+\sum_{s\in\Omega_l}T_{wait}\right) 
\end{align*}
\begin{alignat}{1}
\text{s.t.} \quad & L(s+1,j,k)\leq C(s,j,k), \forall s\in\Omega_l     \\
    &  L(s,j,k) \leq L(s,j,k+1), \forall k\in\{1,\cdots,K\} \\
    &T_{wait} = C(s,j,k)-C(s-1,j,k)-B_{j,k} T_{comp}(k), 
\end{alignat}
\end{subequations}
where if $B_{j,k}>0$ holds, $\sigma(B_{j,k}>0) = 1$, otherwise $\sigma(B_{j,k}>0) = 0$. Constraint (6a) ensures that computation begins only after the loading of the $s$-th quantized expert is complete. Constraint (6b) stipulates that each quantized expert should be loaded sequentially by increasing bit-width, thereby maximizing the reuse of experts with lower bit-width. Constraint (6c) describes how the $s$-th quantized expert waits in the queue until the I/O queue has finished loading.

\textbf{Solution.} Although the above problem can be solved using integer linear programming or dynamic programming, doing so online for every token at each expert layer introduces substantial inference delays. 
To address this, we propose the HEBF algorithm, which prioritizes I/O and computation for experts with higher activation frequencies. Frequently activated experts typically have longer computation times, allowing their execution to overlap with subsequent expert loading, thereby minimizing idle periods. The algorithm proceeds as follows:
1. Construct a queue $\mathcal{Q}_i$ for each expert, sorted in ascending order of bit-width;
2. Pop the bit-width from the "head" of all expert queues and enqueue the element with the highest frequency into the I/O queue;
3. Sequentially load bit-width experts from the I/O queue and begin computation upon completion of loading.
The HEBF algorithm satisfies key constraints: it prioritizes low bit-width experts first (Constraint (6a)), minimizes waiting time by overlapping computation with loading (Constraint (6b)), and ensures that loading completes before computation begins (Constraint (6c)).

\section{IMPLEMENTATION}
We have fully implemented a prototype system of D$^2$MoE with over 2,500 LOC in Python and CUDA in total atop PyTorch.
We use PyTorch’s \emph{triton} library \cite{triton} for I/O-compute parallel programming, and our CUDA programming is based on NVIDIA Ampere and Ada Lovelace architecture. Our approach focuses on the general process of data loading and MoE-based LLM inference, making it easily adaptable to other frameworks, such as TensorRT\cite{tensorRT} and vLLM \cite{2024vllm}.

\section{EVALUATION}
\subsection{Experimental Setup}

\textbf{Models and Datasets.} We evaluate D$^2$MoE through two popular decoder-only MoE-based sparse LLMs: LLaMA-MoE-3.5B~\cite{llama-moe} and Mixtral 8$\times$7B~\cite{LLM_mixtural} have 8 experts per layer and  utilize Top-2 routing, meaning that 2 experts are activated per layer during inference. The pre-trained weights for these models were directly obtained from Hugging Face. Moreover, we use C4 dataset~\cite{C4_dataset} as the training data consists of 2048 random 2048 token segments to train bit-width routers and the calibration data consists of 128 random 2048 token segments to implement MWQ.

\noindent\textbf{Metrics.} We focus on model accuracy and throughput under different memory budget for D$^2$MoE and the baselines. To evaluate model accuracy, we assess language generation performance by reporting perplexity on WikiText2~\cite{WikiText2}, and measure zero-shot performance on several popular benchmarks, including PIQA~\cite{PIQA}, ARC~\cite{arc}, BoolQ~\cite{boolq}, HellaSwag~\cite{hellaswag}, and Winogrande~\cite{WinoGrande}, using the lm-evaluation-harness~\cite{eval-harness}. For throughput, both the input and output lengths are 128.

\noindent\textbf{Hardware.}  We evaluate D$^2$MoE on two prominent kinds of edge devices, as shown in the Table \ref{tab:hardware-env}. The offline preprocessing phase of D$^2$MoE, including fine-tuning the bit-width routers and MWQ, is carried out on a GPU server equipped with NVIDIA RTX 2$\times$A6000.

\begin{table}[t]
\centering
\caption{Hardware environments for evaluation.}
\label{tab:hardware-env}
\resizebox{\linewidth}{!}{ 
\begin{tabular}{l|@{}c@{}c|@{}c@{}c}
\noalign{\hrule height 1.2pt} 
\multirow{2}{*}{\bfseries Hardware} 
  & \multicolumn{2}{c|}{\bfseries Environment 1} 
  & \multicolumn{2}{c}{\bfseries Environment 2} \\
\cline{2-5}
 & \bfseries Device & \bfseries Memory 
 & \bfseries Device & \bfseries Memory \\
\noalign{\hrule height 1.2pt}  
\bfseries GPU
  & NVIDIA RTX 3060
  & 6GB
  & Jetson AGX Orin
  & \multirow{2}{*}{\centering\begin{tabular}{c}64GB\\(SoC share)\end{tabular}} \\
\bfseries CPU
  & Intel Core i7-11800H
  & 32GB
  & ARM Cortex-A78AE
  & \\
  \bfseries Disk
  & Samsung 970 EVO
  & 1T
  & Samsung 970 EVO
  & 1T\\
\bfseries Disk Read
  & \multicolumn{2}{c|}{3.5~GB/s}
  & \multicolumn{2}{c}{3.5~GB/s} \\
\noalign{\hrule height 1.2pt}  
\end{tabular}}
\end{table}

\noindent\textbf{Baselines.} We compare D$^2$MoE with two baselines and three state-of-the-art on-device MoE-based LLM inference frameworks: (1) \emph{Hold-in-Memory}: This method quantizes all experts to INT8 and assumes that all model weights hold in GPU memory. Since INT8 quantization is nearly lossless in terms of accuracy, we use this method as a baseline for evaluating model accuracy. However, it is not memory-efficient. (2) \emph{Matryoshka-Free}:  This method uses GPTQ~\cite{gptq2022} to quantize all expert into multiple versions INT2/3/4 and employs on-demand loading of experts. This baseline demonstrates the performance enhancements that are attributable to MWQ.
(3) \emph{Hold-in-Memory-AWQ}~\cite{lin2024awq}: This method quantizes all experts to INT4 and keeps all model weights in GPU memory. This baseline provides an efficient benchmark for assessing the overhead of dynamically loading experts in D$^2$MoE.
(4) \emph{EdgeMoE}~\cite{Yi2023EdgeMoEFO}: This approach quantizes experts to different bit-width based on their importance, utilizing a pre-loading mechanism to predict and dynamically load the required experts during inference. (5) \emph{MoQE-DynaIO}: This method quantizes all expert weights to a uniform bit-width (e.g., INT4/INT8) using the MoQE quantization method~\cite{MoQE} and dynamically loads them on demand during inference. Similar to D$^2$MoE, the quantized expert weights in all the above methods must be dynamically converted back to the original FP16 format during inference.

\noindent\textbf{Configuration.} We set the group size as 128 in MWQ and utilize two quantized versions of D$^2$MoE: D$^2$MoE-V1, which is compared with INT4 MoE-based LLMs, and D$^2$MoE-V2, which is compared with INT8 MoE-based LLMs. D$^2$MoE-V1 equipped with $b_1 = 2$ and $b_K = 4$, while D$^2$MoE-V2 using $b_1 = 5$ and $b_K = 8$. Furthermore, in D$^2$MoE-V1, the quantized expert capacity is set to $\{0.3, 0.4, 0.3\}$ ,and in D$^2$MoE-V2, the capacity of each expert is 0.25.
\subsection{End-to-End Results}

\textbf{Model Accuracy.} Table \ref{accuracy_storage} presents the the model accuracy of D$^2$MoE with the baselines in LLaMA-MoE-3.5B and Mixtral 8 $\times$7B. It has been demonstrated that D$^2$MoE and Matryoshka-Free exhibit perplexity and zero-shot performance that closely approximates Hold-in-Memory. This suggests that dynamic bit-width selection can effectively guarantee model accuracy. While EdgeMoE demonstrates comparable perplexity to D$^2$MoE, it exhibits reduced accuracy in specific zero-shot tasks. This disparity can be ascribed to the differing significance attributed to the various markers, with EdgeMoE utilizing a predetermined mixture of bit-width to ascertain the importance of the experts, resulting in diminished accuracy.
In contrast, both MoQE-DynaIO-INT4 and MoQE-DynaIO-INT8 demonstrate weaker performance in perplexity and zero-shot tasks. This is due to the quantization method, which results in accuracy loss.

\begin{table*}[htbp]
\caption{Perplexity accuracy (lower is better) and zero-shot accuracy (higher is better) of D$^2$MoE and the baselines in LLaMA-MoE-3.5B  and Mixtral 8$\times$7B.}\label{accuracy_storage}
\centering
\begin{tabular}{cccccccc}
\toprule
Model&Method&Perplexity $\downarrow$ &PiQA&Arc.e&BoolQ&HellaSwag&Winogrande \\
\midrule
\multirow{8}{*}{LLaMA-MoE-3.5B} 
&Hold-in-Memory&14.55&72.32&48.76&65.56&66.34&61.77 \\ \cline{2-8}
&Matryoshke-Free&14.58&72.29&48.71&65.48&66.28&61.76 \\ \cline{2-8}
&Hold-in-Memory-AWQ&15.89&69.32&46.52&62.54&61.55&57.44 \\ \cline{2-8}
&EdgeMoE&14.78&71.46&47.36&64.43&64.32&59.52\\ \cline{2-8}
&MoQE-DynaIO-INT4&15.66&69.34&46.52&62.58&61.53&57.40 \\
&MoQE-DynaIO-INT8&14.55&72.32&48.74&65.58&66.28&61.71\\ \cline{2-8}
&D$^2$MoE-V1&15.68&69.32&46.52&62.50&64.28&59.52\\
&D$^2$MoE-V2&14.58&72.29&48.72&65.51&66.23&61.71\\
\midrule
\multirow{8}{*}{Mixtral 8$\times$7B} 
&Hold-in-Memory&4.04&82.4&82.6&80.56&84.1&76.5 \\ \cline{2-8}
&Matryoshke-Free&4.28&80.29&80.71&78.48&82.28&74.76 \\ \cline{2-8}
&Hold-in-Memory-AWQ&4.25&80.32&81.52&78.54&83.55&75.44 \\ \cline{2-8}
&EdgeMoE&4.38&78.46&80.36&77.43&82.32&75.52\\ \cline{2-8}
&MoQE-DynaIO-INT4&4.25&80.34&81.52&78.58&83.53&75.40 \\
&MoQE-DynaIO-INT8&4.08&82.32&81.74&79.58&83.28&74.71\\ \cline{2-8}
&D$^2$MoE-V1&4.28&81.32&81.52&78.05&82.88&75.52\\
&D$^2$MoE-V2&4.09&82.29&81.72&79.51&83.23&74.71\\
\bottomrule
\end{tabular}
\end{table*}

\begin{figure}[t]
\centering
\subfigure[Throughput of LLaMA-MoE in Environment 1.]{
\includegraphics[width=\linewidth]{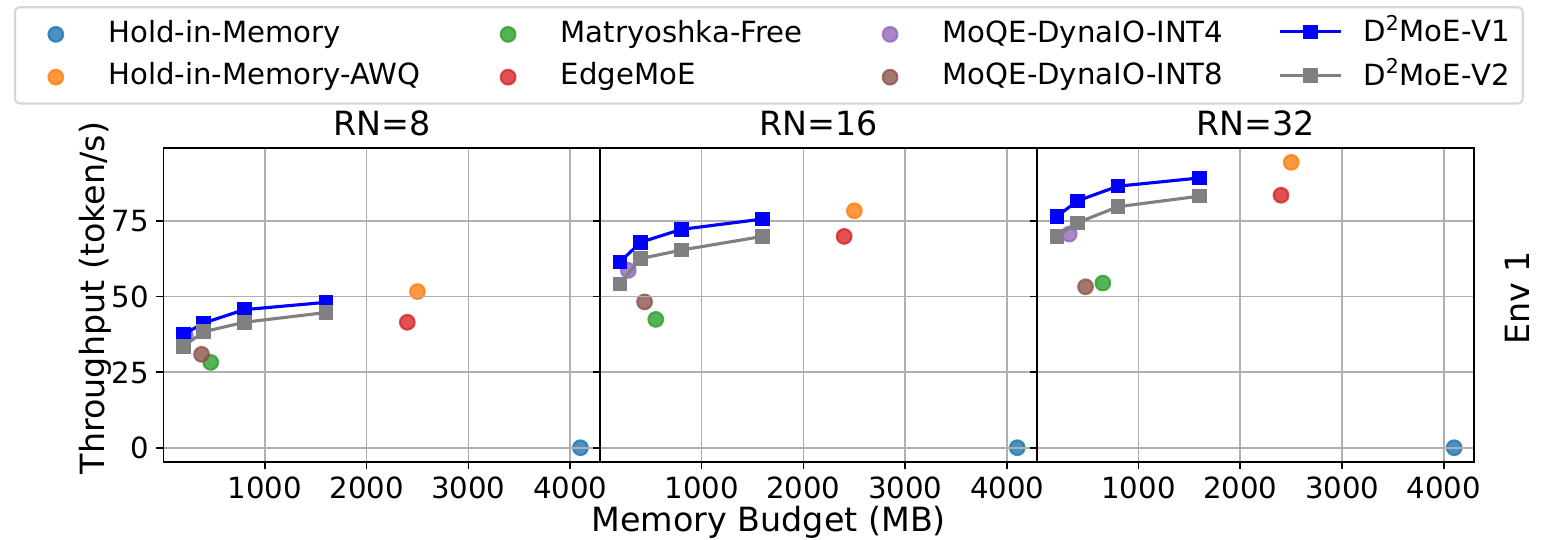} 
}
\subfigure[Throughput of Mixtral 8$\times$7B in Environment 1.]{
\includegraphics[width=0.99\linewidth]{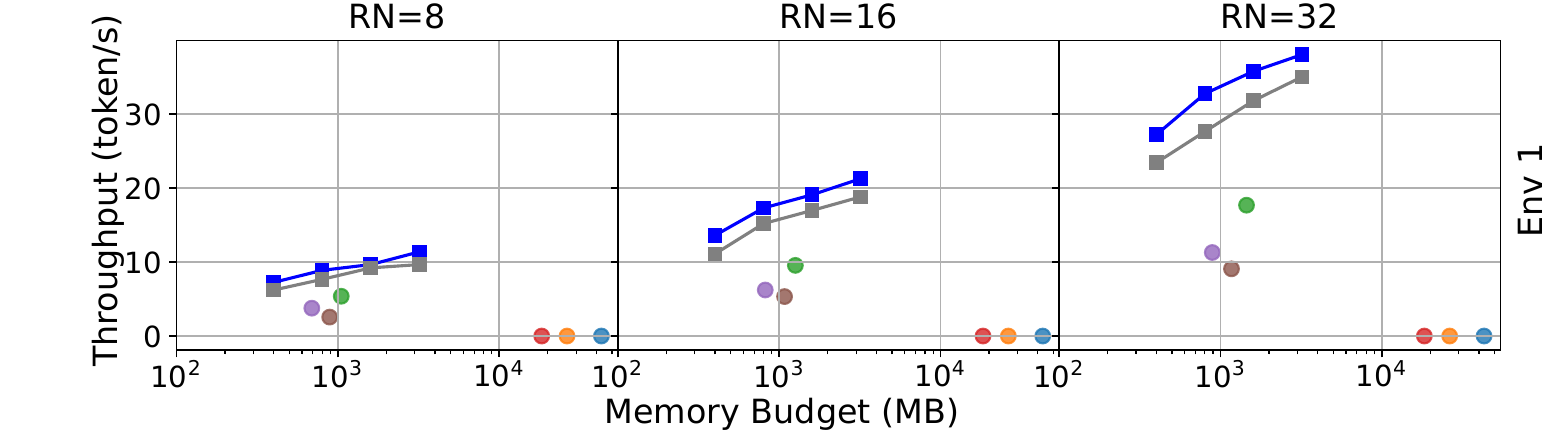} 
}
\subfigure[Throughput of LLaMA-MoE in Environment 2.]{
\includegraphics[width=0.99\linewidth]{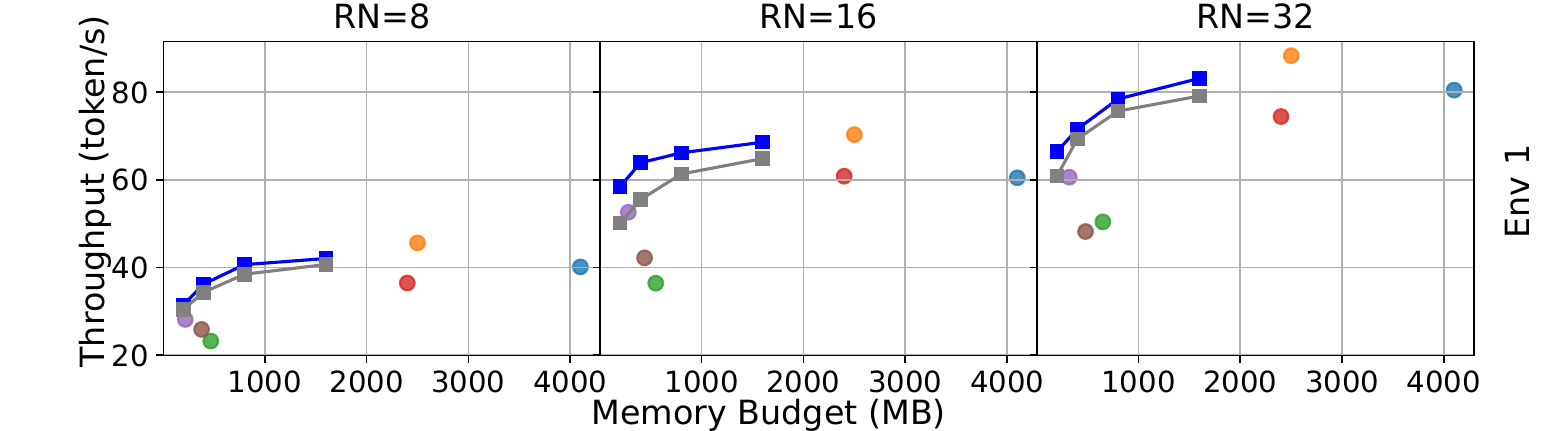}
}
\subfigure[Throughput of Mixtral 8$\times$7B in Environment 2.]{
\includegraphics[width=0.99\linewidth]{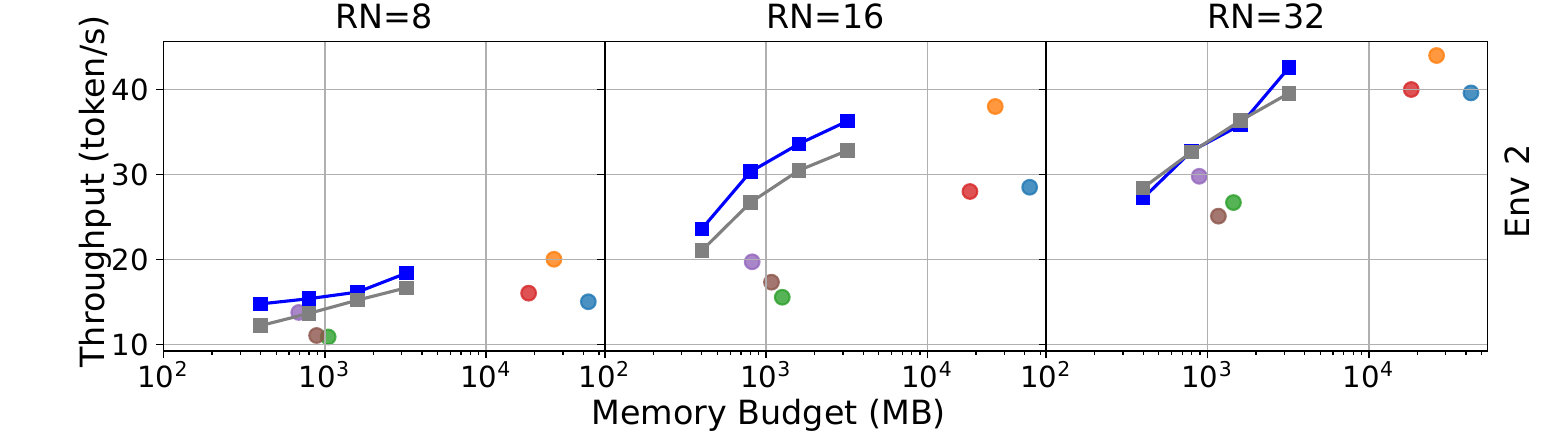}
}
\DeclareGraphicsExtensions.
\caption{Throughput of D$^2$MoE and baselines with different memory budgets.}
\label{throughput}
\end{figure}

\noindent\textbf{Throughput.} Figure \ref{throughput} shows a combined comparison of the throughput of D$^2$MoE with the baseline with different memory budgets in environments 1 and 2. On Mixtral 8$\times$7B, D$^2$MoE achieves a throughput improvement of 1.14$\times$--1.39$\times$ compared to EdgeMoE, and on LLaMA-MoE-3.5B, it improves throughput by 1.06$\times$--1.16$\times$ while reducing memory usage by 33\%--53\%. Against MoQE-DynaIO, D$^2$MoE delivers a 1.42$\times$--3.37$\times$ throughput gain, particularly in Environment 1, underscoring its suitability for memory-constrained edge devices. The throughput of D$^2$MoE approaches that of Hold-in-Memory-AWQ as the memory budget increases, demonstrating near-complete overlap between expert loading and computation. For example, as shown in Figure \ref{throughput}(a) in Environment 1, Hold-in-Memory-AWQ achieves 94.3 tokens/s with 2500MB memory, while D$^2$MoE reaches 89.14 tokens/s with a reduced budget of 1600MB.

D$^2$MoE can adapt to diverse memory budgets on edge devices. For instance, when running Mixtral 8$\times$7B in Environment 1, EdgeMoE and Hold-in-Memory-AWQ fail due to limited GPU memory, while D$^2$MoE achieves a throughput of 38.07 tokens/s. Moreover, D$^2$MoE scales throughput with memory budgets. As shown in Figure \ref{throughput}(c), with 32 requests, throughput rises from 66.45 tokens/s at $M = 200$MB to 83.14 tokens/s at $M = 1600$MB.

\noindent\textbf{Dense LLM Architecture.} We extend D$^2$MoE to dense LLM architecture to compare throughput and peak memory footprint with the traditional approach using a fixed bit-width INT4 on environment 1 with $M = 1600$MB. As illustrated in Figure 11, D$^2$MoE consistently outperforms the method that dynamically loads a fixed bit-width FFN layer with varying request numbers, achieving up to a 1.22$\times$ increase in throughput and up to a 12\% reduction in peak memory consumption. This advantage arises because conventional loading-based quantization methods only load INT4 bit-width, whereas D$^2$MoE can dynamically load bit-width lower than INT4, thereby reducing data transfer overhead more effectively.

Moreover, as the number of requests increases, the  throughput generally exhibits a near-linear growth. However, once the request number reaches 25, the increase slows down due to hardware computational constraints. This phenomenon also exists in MoE-based models. Nevertheless, the performance gains offered by D$^2$MoE are not as pronounced as those observed in MoE-based LLMs, since the FFN layer typically constitutes only 50\%--60\% of the total parameters in dense models. After quantization, this proportion becomes even smaller, thereby shifting the memory bottleneck to the attention layer.

\begin{figure}[t]
  \centering
  \includegraphics[width=\linewidth]{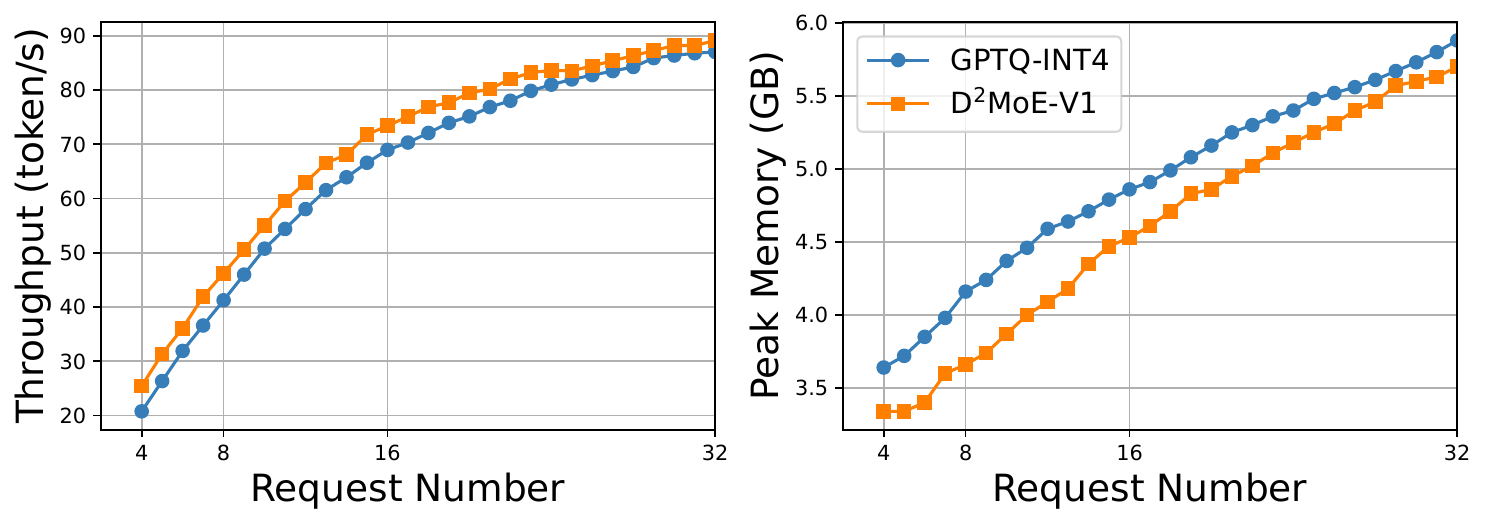}
  \caption{Comparison of GPTQ and D$^2$MoE throughput and peak memory in LLaMA2-13B.}\label{non-MoE}
\end{figure}
\subsection{System Overhead}

\textbf{D$^2$MoE Setup Overhead.} The overhead of setting up D$^2$MoE before inference, including fine-tuning bit-width routers and applying MWQ to quantize experts, is as follows. For LLaMA-MoE-3.5B, fine-tuning the bit-width routers with a batch size of 64 required approximately 2 hours, while MWQ with a batch size of 16 was completed in 10 minutes. For Mixtral 8$\times$7B, fine-tuning with a batch size of 16 took over 4 hours, and MWQ with a batch size of 4 was completed in 20 minutes.

\noindent\textbf{Bit-Width Routing Overhead.} As shown in Table~\ref{bit router overhead}, we evaluate the end-to-end overhead of the D$^2$MoE bit-width router in terms of computation, memory usage, and latency on the LLaMA-MoE-3.5B and Mixtral $8\times 7\text{B}$ models under Environment 1. The additional computation and memory overhead compared to the original MoE-based model is under 0.5\%, with an extra latency of approximately 1.5\%, primarily attributed to the softmax operation in the router. Despite this minor overhead, the bit-width router significantly reduces data transfer by dynamically selecting lower-bit-width experts while maintaining accuracy.

 \begin{table}[htbp]
 \centering
  \caption{Overhead of D$^2$MoE bit-width router on LLaMA-MoE-3.5B and Mixtral 8$\times$7B.}
  \label{bit router overhead}
  \begin{tabular}{cccc}
    \toprule
   Model & Computation & Memory & Latency  \\
    \midrule
LLaMA-MoE-3.5B& 0.28\% & 0.53\% & 1.67\% \\
Mixtral 8$\times$7B& 0.22\% & 0.12\% & 1.04\% \\
    \bottomrule
  \end{tabular}
\end{table}

\begin{figure}[t]
\centering
\subfigure[LLaMA-MoE-3.5B in Environment 1.]{
            \centering
		\includegraphics[width=\linewidth]{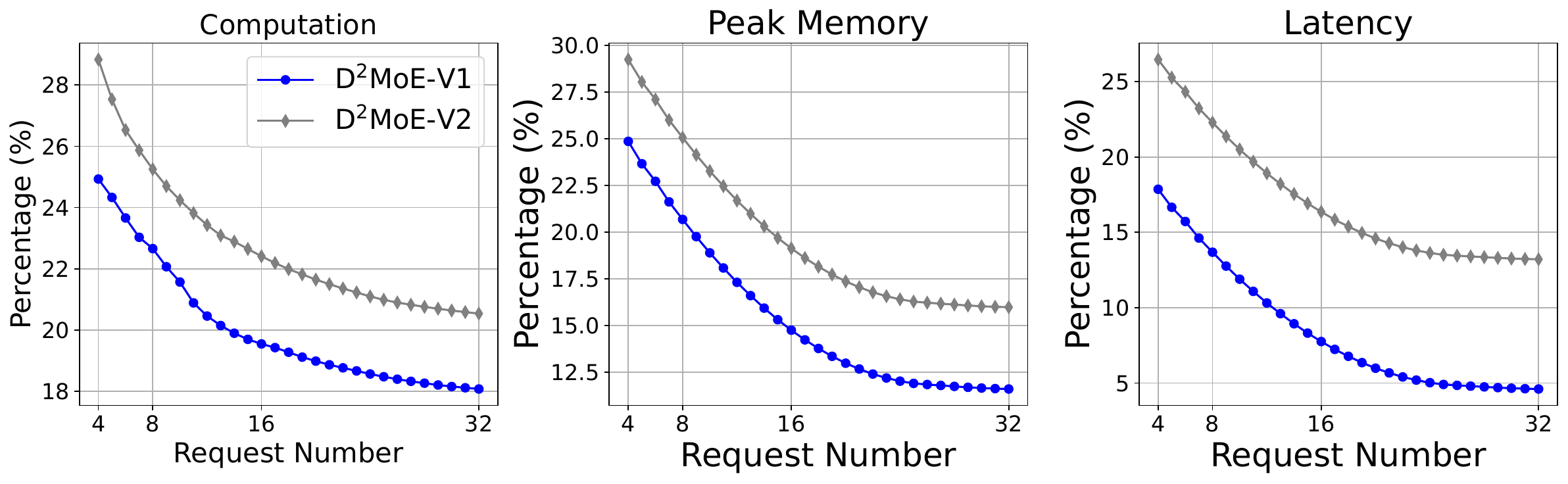}}\\
\subfigure[Mixtral 8$\times$7B in Environment 1]{
            \centering
		\includegraphics[width=\linewidth]{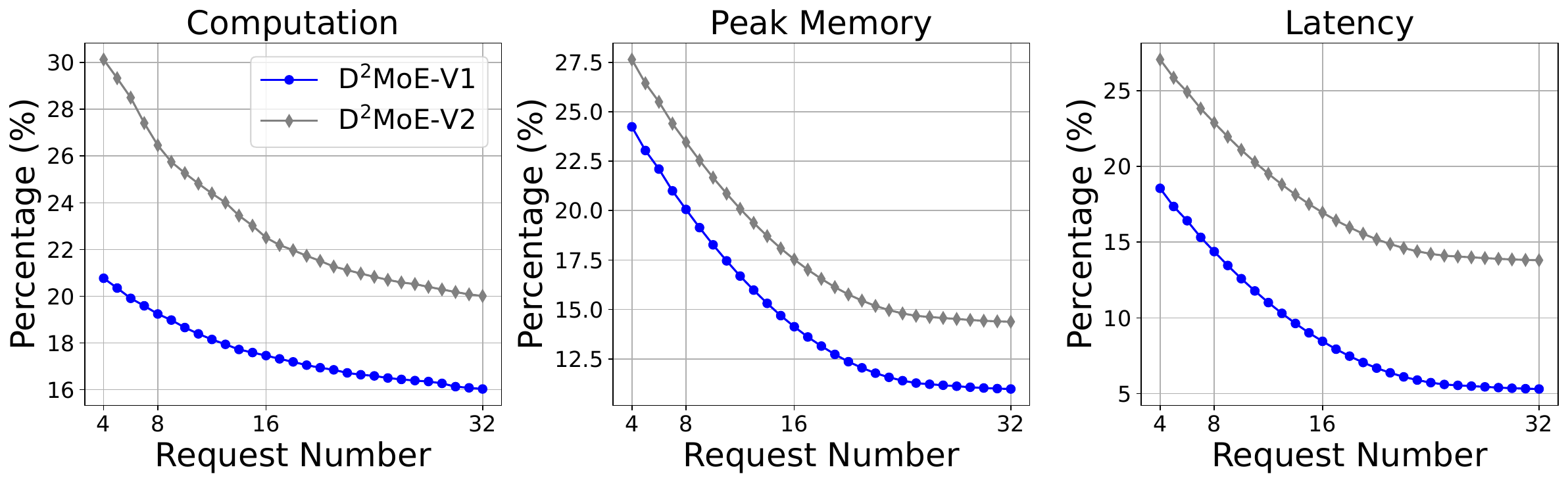}}
\caption{MWQ dequantization overhead of D$^2$MoE.}\label{dequantization overhead}
\end{figure}

\noindent\textbf{MWQ Dequantization Overhead.} As shown in Figure \ref{dequantization overhead}, we evaluated the dequantization overhead of D$^2$MoE in terms of computation, peak memory usage, and latency during inference on the LLaMA-MoE-3.5B and Mixtral $8\times 7\text{B}$ models in Environment 1. Dequantization introduces overhead by converting integer weights to floating-point representations via shift operations. This overhead is more significant with fewer requests, as fewer experts reuse the same bit-width. However, as the number of requests increases, the overhead decreases due to improved weight utilization. For instance, on the Mixtral $8\times 7\text{B}$, when the number of requests increases from 4 to 32, the computational and latency overhead of D$^2$MoE-V1 decreases from 20.77\% and 18.56\% to 16.77\% and 5.3\%, respectively. While FP16 weights temporarily increase peak memory during dequantization, this memory is released immediately after use, resulting in minimal impact on overall inference efficiency.

\begin{figure}[t]
  \centering
  \includegraphics[width=\linewidth]{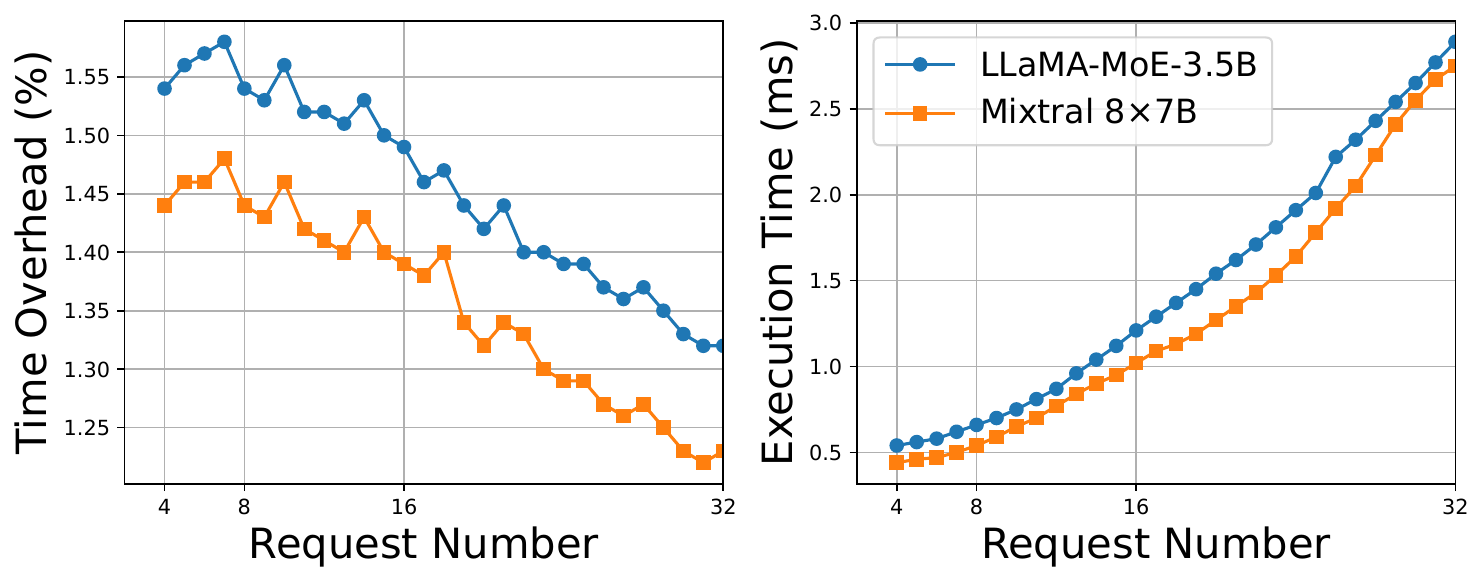}
  \caption{The overall execution time (ms) and the overhead proportion (\%) under different request numbers.}\label{planning overhead}
\end{figure}

\noindent\textbf{Parallelism Planning Overhead.} 
Parallelism planning for various quantized experts constitutes a significant overhead in the D$^2$MoE framework. We conducted profiling on LLaMA-MoE-3.5B and Mixtral 8$\times$7B with request numbers ranging from 4 to 32. The total execution times and the proportion of planning overhead for these models in environment 1 are depicted in Figure \ref{planning overhead}. It is evident that while the execution time for parallelism planning increases with the number of requests, its relative share of the overall inference process decreases. This reduction occurs because the number of loaded quantized experts rises with the increase in requests. Consequently, the additional overhead from parallelism planning remains manageable under on-device inference conditions with multiple requests.

\subsection{Ablation Study}
\begin{figure}[t]
  \centering
  \includegraphics[width=\linewidth]{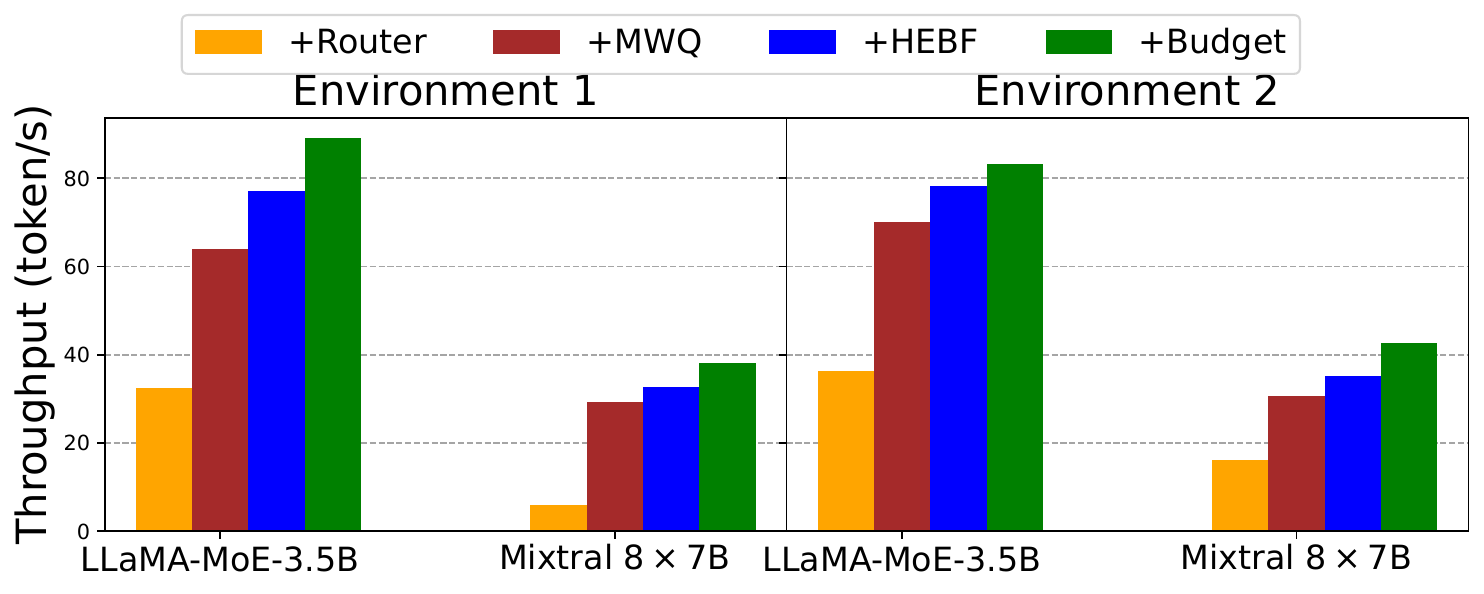}
  \caption{Ablation study for each component of D$^2$MoE on LLaMA-MoE and Mixtral 8$\times$7B.}\label{ablation_study}
\end{figure}

Figure \ref{ablation_study} illustrates the contribution of each D$^2$MoE component to throughput improvement through step-by-step integration. First, token-adaptive bit-width selection ("+Router") is added, enabling dynamic bit-width selection for each expert but still relying on conventional quantization, which stores multiple bit-width versions. Building on +Router, MWQ ("+MWQ") is introduced, nesting lower bit-width within higher ones to reduce expert loading overhead, improving throughput by 1.91$\times$--4.95$\times$. Next, the hottest-expert-bit-first criterion ("+HEBF") is added, parallelizing expert I/O and computation, further reducing I/O-compute bubbles and increasing throughput by 1.11$\times$--1.21$\times$. Finally, integrating expert memory budget ("+Budget") retains frequently activated low bit-width experts in GPU memory, reducing repeated loading and achieving an additional 1.06$\times$--1.21$\times$ improvement. These enhancements due to reduced weight loading and a fine-grained balance of I/O and computation overhead.

\section{DISCUSSION}
The transition of LLMs from a dense to a MoE structure effectively sparsifies the model, reducing computational overhead. Building on this, we further slice the experts at the bit-width level, making the model even more sparse. In the future, we could explore finer granularities, such as sparsification at the neuron level within experts, which remains an open challenge. Currently, the D$^2$MoE system has several limitations:

\textbf{Asynchronous Execution on Edge Devices.} D$^2$MoE does not accommodate a parallel execution strategy for multiple, asynchronously received requests on edge devices. In scenarios where multiple requests initiate reasoning asynchronously, different requests activate distinct layers of experts, substantially increasing bandwidth pressure on edge devices. Developing a more sophisticated quantized expert scheduling algorithm could enhance inference speed by improving the I/O efficiency of quantized experts.

\textbf{Expert Loading Strategy.} D$^2$MoE relies on the on-demand loading of experts, without considering preloading. This approach, while responsive, may introduce additional I/O wait times. Anticipating and preloading the experts likely to be activated later could foster a more efficient I/O/Compute parallel strategy. However, preloading complicates the I/O and Compute strategy, making it more dynamic. We aim to further investigate the viability of this intricate strategy in future research.

\textbf{Suitability for Edge Devices.} D$^2$MoE may not perform optimally on edge devices with limited computational resources, such as smartphones that depend on mobile GPU capabilities. For mobile devices utilizing NPUs, the efficiency of the dequantization process in D$^2$MoE cannot be assured. Nonetheless, these challenges can be solved through tailored NPU arithmetic and system optimizations, areas we plan to explore in our forthcoming efforts.

\section{RELATED WORK}

\textbf{On-device MoE-based LLM Inference.}
PowerInfer \cite{song2023powerinfer} leverages activation sparsity to facilitate inference acceleration on consumer-grade GPUs. EdgeMoE \cite{Yi2023EdgeMoEFO} enhances both memory and computation efficiency by structuring the MoE models within a hierarchical storage framework. AdapMoE \cite{adapMoE} features adaptive expert gating and management by a cohesive algorithm-system co-design, which boosts the inference speeds on edge devices. Fiddler \cite{fiddler} utilizes hybrid GPU-CPU computation on the experts of MoE models, thus minimizing the data transfer between CPUs and GPUs.

\noindent\textbf{LLM Quantization.}
Post-training quantization currently stands as a prevalent technique for compressing LLMs, significantly reducing storage and I/O overhead by transforming high-precision floating-point numbers into low-precision integers. On the one hand, GPTQ \cite{gptq2022}, AWQ \cite{lin2023awq}, and Quant-LLM \cite{quantllm} exemplify weight-only quantization methods, where weights are dynamically converted to the same data type as activation during inference. On the other hand,  weight-activation quantization methods such as SmoothQuant \cite{xiao2023smoothquant}, QServe \cite{qserve}, and AffineQuant \cite{ma2024affinequant} enhance inference speed by directly leveraging integer matrix multiplication operators embedded within hardware.

\noindent\textbf{Pipeline Parallelism for LLM.}
Pipelining techniques are extensively employed to accelerate LLM reasoning by minimizing the gap between I/O and computation. STI \cite{STI_ASPLOS2023} enhances I/O and compute resource utilization by applying mixed bit-width quantization to BERT model weights, thereby accommodating various target latencies and accuracy levels. OTAS \cite{chen2024otaselastictransformerserving} guarantees resilient transformer model services and handles fluctuations in both user requests and query loads through efficient token management.

\section{CONCLUSION}
Aiming at enabling efficient on-device MoE-based LLM serving, we conduct an algorithm-system co-design and propose the D$^2$MoE framework, which introduces adaptive nested quantization based on token properties during multi-request inference. 
D$^2$MoE achieves optimal expert bit-width selection through dual routing and dynamic quantized expert scheduling with minimal I/O-compute parallelism bubbles. 
Realistic evaluation shows that D$^2$MoE significantly improves the latency-memory trade-off with an affordable inference overhead and guaranteed accuracy on edge devices.

\section{ACKNOWLEDGMENTS}
This research was supported by fundings from the Hong Kong RGC General Research Fund (152169/22E, 152228/23E, 162161/24E), Research Impact Fund (No. R5060-19, No. R5011-23), NSFC/RGC Collaborative Research Scheme (Grant No. 62461160332 \& CRS\_HKUST602/24), Collaborative Research Fund (No. C1042-23GF), Areas of Excellence Scheme (AoE/E-601/22-R), and the InnoHK (HKGAI). We also thank MetaX Technology for supporting this research.

\bibliographystyle{ACM-Reference-Format}
\bibliography{sample-base}


\begin{thebibliography}{46}


\ifx \showCODEN    \undefined \def \showCODEN     #1{\unskip}     \fi
\ifx \showDOI      \undefined \def \showDOI       #1{#1}\fi
\ifx \showISBNx    \undefined \def \showISBNx     #1{\unskip}     \fi
\ifx \showISBNxiii \undefined \def \showISBNxiii  #1{\unskip}     \fi
\ifx \showISSN     \undefined \def \showISSN      #1{\unskip}     \fi
\ifx \showLCCN     \undefined \def \showLCCN      #1{\unskip}     \fi
\ifx \shownote     \undefined \def \shownote      #1{#1}          \fi
\ifx \showarticletitle \undefined \def \showarticletitle #1{#1}   \fi
\ifx \showURL      \undefined \def \showURL       {\relax}        \fi
\providecommand\bibfield[2]{#2}
\providecommand\bibinfo[2]{#2}
\providecommand\natexlab[1]{#1}
\providecommand\showeprint[2][]{arXiv:#2}

\bibitem[Bisk et~al\mbox{.}(2020)]%
        {PIQA}
\bibfield{author}{\bibinfo{person}{Yonatan Bisk}, \bibinfo{person}{Rowan Zellers}, \bibinfo{person}{Ronan~Le Bras}, \bibinfo{person}{Jianfeng Gao}, {and} \bibinfo{person}{Yejin Choi}.} \bibinfo{year}{2020}\natexlab{}.
\newblock \showarticletitle{PIQA: Reasoning about Physical Commonsense in Natural Language}. In \bibinfo{booktitle}{\emph{Proceedings of the AAAI conference on artificial intelligence}} \emph{(\bibinfo{series}{AAAI '22}, Vol.~\bibinfo{volume}{34})}. \bibinfo{pages}{7432--7439}.
\newblock


\bibitem[Brown et~al\mbox{.}(2020)]%
        {LLM_task_1}
\bibfield{author}{\bibinfo{person}{Tom~B. Brown}, \bibinfo{person}{Benjamin Mann}, \bibinfo{person}{Nick Ryder}, \bibinfo{person}{Melanie Subbiah}, \bibinfo{person}{Jared Kaplan}, \bibinfo{person}{Prafulla Dhariwal}, \bibinfo{person}{Arvind Neelakantan}, \bibinfo{person}{Pranav Shyam}, \bibinfo{person}{Girish Sastry}, \bibinfo{person}{Amanda Askell}, \bibinfo{person}{Sandhini Agarwal}, \bibinfo{person}{Ariel Herbert-Voss}, \bibinfo{person}{Gretchen Krueger}, \bibinfo{person}{Tom Henighan}, \bibinfo{person}{Rewon Child}, \bibinfo{person}{Aditya Ramesh}, \bibinfo{person}{Daniel~M. Ziegler}, \bibinfo{person}{Jeffrey Wu}, \bibinfo{person}{Clemens Winter}, \bibinfo{person}{Christopher Hesse}, \bibinfo{person}{Mark Chen}, \bibinfo{person}{Eric Sigler}, \bibinfo{person}{Mateusz Litwin}, \bibinfo{person}{Scott Gray}, \bibinfo{person}{Benjamin Chess}, \bibinfo{person}{Jack Clark}, \bibinfo{person}{Christopher Berner}, \bibinfo{person}{Sam McCandlish}, \bibinfo{person}{Alec Radford}, \bibinfo{person}{Ilya Sutskever},
  {and} \bibinfo{person}{Dario Amodei}.} \bibinfo{year}{2020}\natexlab{}.
\newblock \showarticletitle{Language models are few-shot learners}. In \bibinfo{booktitle}{\emph{Proceedings of the 34th International Conference on Neural Information Processing Systems}} \emph{(\bibinfo{series}{NIPS '20})}. \bibinfo{publisher}{Curran Associates Inc.}, Article \bibinfo{articleno}{159}, \bibinfo{numpages}{25}~pages.
\newblock


\bibitem[Cao et~al\mbox{.}(2025)]%
        {cao2024moelightninghighthroughputmoeinference}
\bibfield{author}{\bibinfo{person}{Shiyi Cao}, \bibinfo{person}{Shu Liu}, \bibinfo{person}{Tyler Griggs}, \bibinfo{person}{Peter Schafhalter}, \bibinfo{person}{Xiaoxuan Liu}, \bibinfo{person}{Ying Sheng}, \bibinfo{person}{Joseph~E. Gonzalez}, \bibinfo{person}{Matei Zaharia}, {and} \bibinfo{person}{Ion Stoica}.} \bibinfo{year}{2025}\natexlab{}.
\newblock \bibinfo{title}{MoE-Lightning: High-Throughput MoE Inference on Memory-constrained GPUs}.
\newblock
\newblock


\bibitem[Chen et~al\mbox{.}(2024)]%
        {chen2024otaselastictransformerserving}
\bibfield{author}{\bibinfo{person}{Jinyu Chen}, \bibinfo{person}{Wenchao Xu}, \bibinfo{person}{Zicong Hong}, \bibinfo{person}{Song Guo}, \bibinfo{person}{Haozhao Wang}, \bibinfo{person}{Jie Zhang}, {and} \bibinfo{person}{Deze Zeng}.} \bibinfo{year}{2024}\natexlab{}.
\newblock \showarticletitle{OTAS: An Elastic Transformer Serving System via Token Adaptation}. In \bibinfo{booktitle}{\emph{IEEE Conference on Computer Communications}} \emph{(\bibinfo{series}{INFOCOM '24})}. \bibinfo{pages}{1021--1030}.
\newblock


\bibitem[Chen et~al\mbox{.}(2021)]%
        {codeChen2021}
\bibfield{author}{\bibinfo{person}{Mark Chen}, \bibinfo{person}{Jerry Tworek}, \bibinfo{person}{Heewoo Jun}, \bibinfo{person}{Qiming Yuan}, \bibinfo{person}{Henrique Ponde}, \bibinfo{person}{Jared Kaplan}, \bibinfo{person}{Harrison Edwards}, \bibinfo{person}{Yura Burda}, \bibinfo{person}{Nicholas Joseph}, \bibinfo{person}{Greg Brockman}, \bibinfo{person}{Alex Ray}, \bibinfo{person}{Raul Puri}, \bibinfo{person}{Gretchen Krueger}, \bibinfo{person}{Michael Petrov}, \bibinfo{person}{Heidy Khlaaf}, \bibinfo{person}{Girish Sastry}, \bibinfo{person}{Pamela Mishkin}, \bibinfo{person}{Brooke Chan}, \bibinfo{person}{Scott Gray}, \bibinfo{person}{Nick Ryder}, \bibinfo{person}{Mikhail Pavlov}, \bibinfo{person}{Alethea Power}, \bibinfo{person}{Lukasz Kaiser}, \bibinfo{person}{Mohammad Bavarian}, \bibinfo{person}{Clemens Winter}, \bibinfo{person}{Philippe Tillet}, \bibinfo{person}{Felipe~Petroski Such}, \bibinfo{person}{David~W. Cummings}, \bibinfo{person}{Matthias Plappert}, \bibinfo{person}{Fotios Chantzis},
  \bibinfo{person}{Elizabeth Barnes}, \bibinfo{person}{Ariel Herbert-Voss}, \bibinfo{person}{William~H. Guss}, \bibinfo{person}{Alex Nichol}, \bibinfo{person}{Igor Babuschkin}, \bibinfo{person}{Suchir Balaji}, \bibinfo{person}{Shantanu Jain}, \bibinfo{person}{Andrew Carr}, \bibinfo{person}{Jan Leike}, \bibinfo{person}{Joshua Achiam}, \bibinfo{person}{Vedant Misra}, \bibinfo{person}{Evan Morikawa}, \bibinfo{person}{Alec Radford}, \bibinfo{person}{Matthew~M. Knight}, \bibinfo{person}{Miles Brundage}, \bibinfo{person}{Mira Murati}, \bibinfo{person}{Katie Mayer}, \bibinfo{person}{Peter Welinder}, \bibinfo{person}{Bob McGrew}, \bibinfo{person}{Dario Amodei}, \bibinfo{person}{Sam McCandlish}, \bibinfo{person}{Ilya Sutskever}, {and} \bibinfo{person}{Wojciech Zaremba}.} \bibinfo{year}{2021}\natexlab{}.
\newblock \showarticletitle{Evaluating Large Language Models Trained on Code}.
\newblock \bibinfo{journal}{\emph{ArXiv}}  \bibinfo{volume}{abs/2107.03374} (\bibinfo{year}{2021}).
\newblock


\bibitem[Clark et~al\mbox{.}(2019)]%
        {boolq}
\bibfield{author}{\bibinfo{person}{Christopher Clark}, \bibinfo{person}{Kenton Lee}, \bibinfo{person}{Ming-Wei Chang}, \bibinfo{person}{Tom Kwiatkowski}, \bibinfo{person}{Michael Collins}, {and} \bibinfo{person}{Kristina Toutanova}.} \bibinfo{year}{2019}\natexlab{}.
\newblock \showarticletitle{BoolQ: Exploring the Surprising Difficulty of Natural Yes/No Questions}. In \bibinfo{booktitle}{\emph{Proceedings of the 2019 Conference of the North American Chapter of the Association for Computational Linguistics: Human Language Technologies}}, Vol.~\bibinfo{volume}{1}. \bibinfo{pages}{2924--2936}.
\newblock


\bibitem[Clark et~al\mbox{.}(2018)]%
        {arc}
\bibfield{author}{\bibinfo{person}{Peter Clark}, \bibinfo{person}{Isaac Cowhey}, \bibinfo{person}{Oren Etzioni}, \bibinfo{person}{Tushar Khot}, \bibinfo{person}{Ashish Sabharwal}, \bibinfo{person}{Carissa Schoenick}, {and} \bibinfo{person}{Oyvind Tafjord}.} \bibinfo{year}{2018}\natexlab{}.
\newblock \bibinfo{title}{Think you have Solved Question Answering? Try ARC, the AI2 Reasoning Challenge}.
\newblock
\newblock
\showeprint[arxiv]{1803.05457}


\bibitem[Fedus et~al\mbox{.}(2022)]%
        {switchtransformer}
\bibfield{author}{\bibinfo{person}{William Fedus}, \bibinfo{person}{Barret Zoph}, {and} \bibinfo{person}{Noam Shazeer}.} \bibinfo{year}{2022}\natexlab{}.
\newblock \showarticletitle{Switch transformers: scaling to trillion parameter models with simple and efficient sparsity}.
\newblock \bibinfo{journal}{\emph{The Journal of Machine Learning Research}} \bibinfo{volume}{23}, \bibinfo{number}{1} (\bibinfo{date}{jan} \bibinfo{year}{2022}), \bibinfo{numpages}{39}~pages.
\newblock
\showISSN{1532-4435}


\bibitem[Frantar et~al\mbox{.}(2023)]%
        {gptq2022}
\bibfield{author}{\bibinfo{person}{Elias Frantar}, \bibinfo{person}{Saleh Ashkboos}, \bibinfo{person}{Torsten Hoefler}, {and} \bibinfo{person}{Dan Alistarh}.} \bibinfo{year}{2023}\natexlab{}.
\newblock \showarticletitle{{GPTQ}: Accurate Post-training Compression for Generative Pretrained Transformers}.
\newblock \bibinfo{journal}{\emph{The Eleventh International Conference on Learning Representations}} (\bibinfo{year}{2023}).
\newblock


\bibitem[Gao et~al\mbox{.}(2024)]%
        {eval-harness}
\bibfield{author}{\bibinfo{person}{Leo Gao}, \bibinfo{person}{Jonathan Tow}, \bibinfo{person}{Baber Abbasi}, \bibinfo{person}{Stella Biderman}, \bibinfo{person}{Sid Black}, \bibinfo{person}{Anthony DiPofi}, \bibinfo{person}{Charles Foster}, \bibinfo{person}{Laurence Golding}, \bibinfo{person}{Jeffrey Hsu}, \bibinfo{person}{Alain Le~Noac'h}, \bibinfo{person}{Haonan Li}, \bibinfo{person}{Kyle McDonell}, \bibinfo{person}{Niklas Muennighoff}, \bibinfo{person}{Chris Ociepa}, \bibinfo{person}{Jason Phang}, \bibinfo{person}{Laria Reynolds}, \bibinfo{person}{Hailey Schoelkopf}, \bibinfo{person}{Aviya Skowron}, \bibinfo{person}{Lintang Sutawika}, \bibinfo{person}{Eric Tang}, \bibinfo{person}{Anish Thite}, \bibinfo{person}{Ben Wang}, \bibinfo{person}{Kevin Wang}, {and} \bibinfo{person}{Andy Zou}.} \bibinfo{year}{2024}\natexlab{}.
\newblock \bibinfo{title}{A framework for few-shot language model evaluation}.
\newblock
\newblock


\bibitem[Gerganov(2023)]%
        {LLMmobile_llamacpp}
\bibfield{author}{\bibinfo{person}{Georgi Gerganov}.} \bibinfo{year}{2023}\natexlab{}.
\newblock \bibinfo{booktitle}{\emph{llama.cpp}}.
\newblock
\urldef\tempurl%
\url{https://github.com/ ggerganov/llama.cpp}
\showURL{%
\tempurl}


\bibitem[Github(2022)]%
        {LLM_task_copilot}
\bibfield{author}{\bibinfo{person}{Github}.} \bibinfo{year}{2022}\natexlab{}.
\newblock \bibinfo{booktitle}{\emph{Copilot}}.
\newblock
\urldef\tempurl%
\url{https://github.com/features/ copilot}
\showURL{%
\tempurl}


\bibitem[Gong et~al\mbox{.}(2024)]%
        {gong-etal-2024-mixture}
\bibfield{author}{\bibinfo{person}{Zhuocheng Gong}, \bibinfo{person}{Ang Lv}, \bibinfo{person}{Jian Guan}, \bibinfo{person}{Wei Wu}, \bibinfo{person}{Huishuai Zhang}, \bibinfo{person}{Minlie Huang}, \bibinfo{person}{Dongyan Zhao}, {and} \bibinfo{person}{Rui Yan}.} \bibinfo{year}{2024}\natexlab{}.
\newblock \showarticletitle{Mixture-of-Modules: Reinventing Transformers as Dynamic Assemblies of Modules}. In \bibinfo{booktitle}{\emph{Proceedings of the 2024 Conference on Empirical Methods in Natural Language Processing}}. \bibinfo{pages}{20924--20938}.
\newblock


\bibitem[Guo et~al\mbox{.}(2023)]%
        {STI_ASPLOS2023}
\bibfield{author}{\bibinfo{person}{Liwei Guo}, \bibinfo{person}{Wonkyo Choe}, {and} \bibinfo{person}{Felix~Xiaozhu Lin}.} \bibinfo{year}{2023}\natexlab{}.
\newblock \showarticletitle{STI: Turbocharge NLP Inference at the Edge via Elastic Pipelining}. In \bibinfo{booktitle}{\emph{Proceedings of the 28th ACM International Conference on Architectural Support for Programming Languages and Operating Systems}} \emph{(\bibinfo{series}{ASPLOS 2023})}. \bibinfo{pages}{791–803}.
\newblock


\bibitem[Guo and Lin(2022)]%
        {IOcomp2022}
\bibfield{author}{\bibinfo{person}{Liwei Guo} {and} \bibinfo{person}{Felix~Xiaozhu Lin}.} \bibinfo{year}{2022}\natexlab{}.
\newblock \showarticletitle{Minimum viable device drivers for ARM trustzone}. In \bibinfo{booktitle}{\emph{Proceedings of the Seventeenth European Conference on Computer Systems}} \emph{(\bibinfo{series}{EuroSys '22})}. \bibinfo{pages}{300–316}.
\newblock


\bibitem[Huang et~al\mbox{.}(2025)]%
        {huang2024mc-moe}
\bibfield{author}{\bibinfo{person}{Wei Huang}, \bibinfo{person}{Yue Liao}, \bibinfo{person}{Jianhui Liu}, \bibinfo{person}{Ruifei He}, \bibinfo{person}{Haoru Tan}, \bibinfo{person}{Shiming Zhang}, \bibinfo{person}{Hongsheng Li}, \bibinfo{person}{Si Liu}, {and} \bibinfo{person}{Xiaojuan Qi}.} \bibinfo{year}{2025}\natexlab{}.
\newblock \showarticletitle{Mc-moe: Mixture compressor for mixture-of-experts llms gains more}.
\newblock \bibinfo{journal}{\emph{The Eleventh International Conference on Learning Representations}} (\bibinfo{year}{2025}).
\newblock


\bibitem[Jiang et~al\mbox{.}(2024)]%
        {LLM_mixtural}
\bibfield{author}{\bibinfo{person}{Albert~Q. Jiang}, \bibinfo{person}{Alexandre Sablayrolles}, \bibinfo{person}{Antoine Roux}, \bibinfo{person}{Arthur Mensch}, \bibinfo{person}{Blanche Savary}, \bibinfo{person}{Chris Bamford}, \bibinfo{person}{Devendra~Singh Chaplot}, \bibinfo{person}{Diego de~las Casas}, \bibinfo{person}{Emma~Bou Hanna}, \bibinfo{person}{Florian Bressand}, \bibinfo{person}{Gianna Lengyel}, \bibinfo{person}{Guillaume Bour}, \bibinfo{person}{Guillaume Lample}, \bibinfo{person}{Lélio~Renard Lavaud}, \bibinfo{person}{Lucile Saulnier}, \bibinfo{person}{Marie-Anne Lachaux}, \bibinfo{person}{Pierre Stock}, \bibinfo{person}{Sandeep Subramanian}, \bibinfo{person}{Sophia Yang}, \bibinfo{person}{Szymon Antoniak}, \bibinfo{person}{Teven~Le Scao}, \bibinfo{person}{Théophile Gervet}, \bibinfo{person}{Thibaut Lavril}, \bibinfo{person}{Thomas Wang}, \bibinfo{person}{Timothée Lacroix}, {and} \bibinfo{person}{William~El Sayed}.} \bibinfo{year}{2024}\natexlab{}.
\newblock \bibinfo{title}{Mixtral of Experts}.
\newblock
\newblock
\showeprint[arxiv]{2401.04088}


\bibitem[Kamahori et~al\mbox{.}(2024)]%
        {fiddler}
\bibfield{author}{\bibinfo{person}{Keisuke Kamahori}, \bibinfo{person}{Yile Gu}, \bibinfo{person}{Kan Zhu}, {and} \bibinfo{person}{Baris Kasikci}.} \bibinfo{year}{2024}\natexlab{}.
\newblock \showarticletitle{Fiddler: {CPU}-{GPU} Orchestration for Fast Inference of Mixture-of-Experts Models}. In \bibinfo{booktitle}{\emph{5th Workshop on practical ML for limited/low resource settings}}.
\newblock


\bibitem[Kim et~al\mbox{.}(2023)]%
        {MoQE}
\bibfield{author}{\bibinfo{person}{Young~Jin Kim}, \bibinfo{person}{Raffy Fahim}, {and} \bibinfo{person}{Hany~Hassan Awadalla}.} \bibinfo{year}{2023}\natexlab{}.
\newblock \bibinfo{title}{Mixture of Quantized Experts (MoQE): Complementary Effect of Low-bit Quantization and Robustness}.
\newblock
\newblock
\showeprint[arxiv]{2310.02410}


\bibitem[Kwon et~al\mbox{.}(2023)]%
        {2024vllm}
\bibfield{author}{\bibinfo{person}{Woosuk Kwon}, \bibinfo{person}{Zhuohan Li}, \bibinfo{person}{Siyuan Zhuang}, \bibinfo{person}{Ying Sheng}, \bibinfo{person}{Lianmin Zheng}, \bibinfo{person}{Cody~Hao Yu}, \bibinfo{person}{Joseph Gonzalez}, \bibinfo{person}{Hao Zhang}, {and} \bibinfo{person}{Ion Stoica}.} \bibinfo{year}{2023}\natexlab{}.
\newblock \showarticletitle{Efficient Memory Management for Large Language Model Serving with PagedAttention}. In \bibinfo{booktitle}{\emph{Proceedings of the 29th Symposium on Operating Systems Principles}} \emph{(\bibinfo{series}{SOSP '23})}. \bibinfo{pages}{611–626}.
\newblock


\bibitem[Li et~al\mbox{.}(2024)]%
        {flexnn_mobicom2024}
\bibfield{author}{\bibinfo{person}{Xiangyu Li}, \bibinfo{person}{Yuanchun Li}, \bibinfo{person}{Yuanzhe Li}, \bibinfo{person}{Ting Cao}, {and} \bibinfo{person}{Yunxin Liu}.} \bibinfo{year}{2024}\natexlab{}.
\newblock \showarticletitle{FlexNN: Efficient and Adaptive DNN Inference on Memory-Constrained Edge Devices}. In \bibinfo{booktitle}{\emph{Proceedings of the 30th Annual International Conference on Mobile Computing and Networking}} \emph{(\bibinfo{series}{MobiCom '24})}. \bibinfo{pages}{709–723}.
\newblock


\bibitem[Lin et~al\mbox{.}(2024a)]%
        {lin2023awq}
\bibfield{author}{\bibinfo{person}{Ji Lin}, \bibinfo{person}{Jiaming Tang}, \bibinfo{person}{Haotian Tang}, \bibinfo{person}{Shang Yang}, \bibinfo{person}{Wei-Ming Chen}, \bibinfo{person}{Wei-Chen Wang}, \bibinfo{person}{Guangxuan Xiao}, \bibinfo{person}{Xingyu Dang}, \bibinfo{person}{Chuang Gan}, {and} \bibinfo{person}{Song Han}.} \bibinfo{year}{2024}\natexlab{a}.
\newblock \showarticletitle{AWQ: Activation-aware Weight Quantization for LLM Compression and Acceleration}. In \bibinfo{booktitle}{\emph{Proceedings of Machine Learning and Systems}} \emph{(\bibinfo{series}{MLSys '24}, Vol.~\bibinfo{volume}{6})}. \bibinfo{pages}{87--100}.
\newblock


\bibitem[Lin et~al\mbox{.}(2024b)]%
        {lin2024awq}
\bibfield{author}{\bibinfo{person}{Ji Lin}, \bibinfo{person}{Jiaming Tang}, \bibinfo{person}{Haotian Tang}, \bibinfo{person}{Shang Yang}, \bibinfo{person}{Wei-Ming Chen}, \bibinfo{person}{Wei-Chen Wang}, \bibinfo{person}{Guangxuan Xiao}, \bibinfo{person}{Xingyu Dang}, \bibinfo{person}{Chuang Gan}, {and} \bibinfo{person}{Song Han}.} \bibinfo{year}{2024}\natexlab{b}.
\newblock \showarticletitle{AWQ: Activation-aware Weight Quantization for On-Device LLM Compression and Acceleration}.
\newblock \bibinfo{journal}{\emph{Proceedings of Machine Learning and Systems}}  \bibinfo{volume}{6} (\bibinfo{year}{2024}), \bibinfo{pages}{87--100}.
\newblock


\bibitem[Lin* et~al\mbox{.}(2024)]%
        {qserve}
\bibfield{author}{\bibinfo{person}{Yujun Lin*}, \bibinfo{person}{Haotian Tang*}, \bibinfo{person}{Shang Yang*}, \bibinfo{person}{Zhekai Zhang}, \bibinfo{person}{Guangxuan Xiao}, \bibinfo{person}{Chuang Gan}, {and} \bibinfo{person}{Song Han}.} \bibinfo{year}{2024}\natexlab{}.
\newblock \showarticletitle{QServe: W4A8KV4 Quantization and System Co-design for Efficient LLM Serving}.
\newblock \bibinfo{journal}{\emph{arXiv preprint arXiv:2405.04532}} (\bibinfo{year}{2024}).
\newblock


\bibitem[Ma et~al\mbox{.}(2024)]%
        {ma2024affinequant}
\bibfield{author}{\bibinfo{person}{Yuexiao Ma}, \bibinfo{person}{Huixia Li}, \bibinfo{person}{Xiawu Zheng}, \bibinfo{person}{Feng Ling}, \bibinfo{person}{Xuefeng Xiao}, \bibinfo{person}{Rui Wang}, \bibinfo{person}{Shilei Wen}, \bibinfo{person}{Fei Chao}, {and} \bibinfo{person}{Rongrong Ji}.} \bibinfo{year}{2024}\natexlab{}.
\newblock \showarticletitle{AffineQuant: Affine Transformation Quantization for Large Language Models}. In \bibinfo{booktitle}{\emph{The Twelfth International Conference on Learning Representations}} \emph{(\bibinfo{series}{ICLR '24})}.
\newblock


\bibitem[Merity et~al\mbox{.}(2016)]%
        {WikiText2}
\bibfield{author}{\bibinfo{person}{Stephen Merity}, \bibinfo{person}{Caiming Xiong}, \bibinfo{person}{James Bradbury}, {and} \bibinfo{person}{Richard Socher}.} \bibinfo{year}{2016}\natexlab{}.
\newblock \bibinfo{title}{Pointer Sentinel Mixture Models}.
\newblock
\newblock
\showeprint[arxiv]{1609.07843}


\bibitem[NVIDIA(023b)]%
        {tensorRT}
\bibfield{author}{\bibinfo{person}{NVIDIA}.} \bibinfo{year}{2023b}\natexlab{}.
\newblock \bibinfo{booktitle}{\emph{NVIDIA. Tensorrt-llm}}.
\newblock
\urldef\tempurl%
\url{https://github.com/ NVIDIA/TensorRT-LLM}
\showURL{%
\tempurl}


\bibitem[OpenAI(2022)]%
        {LLM_task_chatgpt}
\bibfield{author}{\bibinfo{person}{OpenAI}.} \bibinfo{year}{2022}\natexlab{}.
\newblock \bibinfo{booktitle}{\emph{Chatgpt}}.
\newblock
\urldef\tempurl%
\url{https://openai.com/blog/ chatgpt}
\showURL{%
\tempurl}


\bibitem[Park et~al\mbox{.}(2024)]%
        {park2024anyprecision}
\bibfield{author}{\bibinfo{person}{Yeonhong Park}, \bibinfo{person}{Jake Hyun}, \bibinfo{person}{SangLyul Cho}, \bibinfo{person}{Bonggeun Sim}, {and} \bibinfo{person}{Jae~W. Lee}.} \bibinfo{year}{2024}\natexlab{}.
\newblock \showarticletitle{Any-Precision LLM: Low-Cost Deployment of Multiple, Different-Sized LLMs}. In \bibinfo{booktitle}{\emph{Proceedings of the 41st International Conference on Machine Learning}} \emph{(\bibinfo{series}{ICML '24})}.
\newblock


\bibitem[Raffel et~al\mbox{.}(2020)]%
        {C4_dataset}
\bibfield{author}{\bibinfo{person}{Colin Raffel}, \bibinfo{person}{Noam Shazeer}, \bibinfo{person}{Adam Roberts}, \bibinfo{person}{Katherine Lee}, \bibinfo{person}{Sharan Narang}, \bibinfo{person}{Michael Matena}, \bibinfo{person}{Yanqi Zhou}, \bibinfo{person}{Wei Li}, {and} \bibinfo{person}{Peter~J. Liu}.} \bibinfo{year}{2020}\natexlab{}.
\newblock \showarticletitle{Exploring the limits of transfer learning with a unified text-to-text transformer}.
\newblock \bibinfo{journal}{\emph{Journal of Machine Learning Research}} \bibinfo{volume}{21}, \bibinfo{number}{1}, Article \bibinfo{articleno}{140} (\bibinfo{year}{2020}), \bibinfo{numpages}{67}~pages.
\newblock


\bibitem[Raposo et~al\mbox{.}(2024a)]%
        {raposo2024mod}
\bibfield{author}{\bibinfo{person}{David Raposo}, \bibinfo{person}{Sam Ritter}, \bibinfo{person}{Blake Richards}, \bibinfo{person}{Timothy Lillicrap}, \bibinfo{person}{Peter~Conway Humphreys}, {and} \bibinfo{person}{Adam Santoro}.} \bibinfo{year}{2024}\natexlab{a}.
\newblock \bibinfo{title}{Mixture-of-Depths: Dynamically allocating compute in transformer-based language models}.
\newblock
\newblock
\showeprint[arxiv]{2404.02258}


\bibitem[Raposo et~al\mbox{.}(2024b)]%
        {2024mixtureofdepths}
\bibfield{author}{\bibinfo{person}{David Raposo}, \bibinfo{person}{Sam Ritter}, \bibinfo{person}{Blake Richards}, \bibinfo{person}{Timothy Lillicrap}, \bibinfo{person}{Peter~Conway Humphreys}, {and} \bibinfo{person}{Adam Santoro}.} \bibinfo{year}{2024}\natexlab{b}.
\newblock \bibinfo{title}{Mixture-of-Depths: Dynamically allocating compute in transformer-based language models}.
\newblock
\newblock
\showeprint[arxiv]{2404.02258}


\bibitem[Sakaguchi et~al\mbox{.}(2021)]%
        {WinoGrande}
\bibfield{author}{\bibinfo{person}{Keisuke Sakaguchi}, \bibinfo{person}{Ronan~Le Bras}, \bibinfo{person}{Chandra Bhagavatula}, {and} \bibinfo{person}{Yejin Choi}.} \bibinfo{year}{2021}\natexlab{}.
\newblock \showarticletitle{WinoGrande: an adversarial winograd schema challenge at scale}.
\newblock \bibinfo{journal}{\emph{Commun. ACM}} \bibinfo{volume}{64}, \bibinfo{number}{9} (\bibinfo{date}{aug} \bibinfo{year}{2021}), \bibinfo{pages}{99–106}.
\newblock


\bibitem[Song et~al\mbox{.}(2023)]%
        {song2023powerinfer}
\bibfield{author}{\bibinfo{person}{Yixin Song}, \bibinfo{person}{Zeyu Mi}, \bibinfo{person}{Haotong Xie}, {and} \bibinfo{person}{Haibo Chen}.} \bibinfo{year}{2023}\natexlab{}.
\newblock \bibinfo{title}{PowerInfer: Fast Large Language Model Serving with a Consumer-grade GPU}.
\newblock
\newblock
\showeprint[arxiv]{2312.12456}


\bibitem[Sun et~al\mbox{.}(2024)]%
        {sun2024llmserve}
\bibfield{author}{\bibinfo{person}{Biao Sun}, \bibinfo{person}{Ziming Huang}, \bibinfo{person}{Hanyu Zhao}, \bibinfo{person}{Wencong Xiao}, \bibinfo{person}{Xinyi Zhang}, \bibinfo{person}{Yong Li}, {and} \bibinfo{person}{Wei Lin}.} \bibinfo{year}{2024}\natexlab{}.
\newblock \showarticletitle{Llumnix: Dynamic Scheduling for Large Language Model Serving}.
\newblock \bibinfo{journal}{\emph{18th USENIX Symposium on Operating Systems Design and Implementation}} (\bibinfo{year}{2024}).
\newblock


\bibitem[Tillet et~al\mbox{.}(2019)]%
        {triton}
\bibfield{author}{\bibinfo{person}{Philippe Tillet}, \bibinfo{person}{H.~T. Kung}, {and} \bibinfo{person}{David Cox}.} \bibinfo{year}{2019}\natexlab{}.
\newblock \showarticletitle{Triton: an intermediate language and compiler for tiled neural network computations}. In \bibinfo{booktitle}{\emph{Proceedings of the 3rd ACM SIGPLAN International Workshop on Machine Learning and Programming Languages}} \emph{(\bibinfo{series}{MAPL '19})}. \bibinfo{pages}{10–19}.
\newblock


\bibitem[Touvron et~al\mbox{.}(2023)]%
        {llama2}
\bibfield{author}{\bibinfo{person}{Hugo Touvron}, \bibinfo{person}{Louis Martin}, \bibinfo{person}{Kevin Stone}, \bibinfo{person}{Peter Albert}, \bibinfo{person}{Amjad Almahairi}, \bibinfo{person}{Yasmine Babaei}, \bibinfo{person}{Nikolay Bashlykov}, \bibinfo{person}{Soumya Batra}, \bibinfo{person}{Prajjwal Bhargava}, \bibinfo{person}{Shruti Bhosale}, \bibinfo{person}{Dan Bikel}, \bibinfo{person}{Lukas Blecher}, \bibinfo{person}{Cristian~Canton Ferrer}, \bibinfo{person}{Moya Chen}, \bibinfo{person}{Guillem Cucurull}, \bibinfo{person}{David Esiobu}, \bibinfo{person}{Jude Fernandes}, \bibinfo{person}{Jeremy Fu}, \bibinfo{person}{Wenyin Fu}, \bibinfo{person}{Brian Fuller}, \bibinfo{person}{Cynthia Gao}, \bibinfo{person}{Vedanuj Goswami}, \bibinfo{person}{Naman Goyal}, \bibinfo{person}{Anthony Hartshorn}, \bibinfo{person}{Saghar Hosseini}, \bibinfo{person}{Rui Hou}, \bibinfo{person}{Hakan Inan}, \bibinfo{person}{Marcin Kardas}, \bibinfo{person}{Viktor Kerkez}, \bibinfo{person}{Madian Khabsa},
  \bibinfo{person}{Isabel Kloumann}, \bibinfo{person}{Artem Korenev}, \bibinfo{person}{Punit~Singh Koura}, \bibinfo{person}{Marie-Anne Lachaux}, \bibinfo{person}{Thibaut Lavril}, \bibinfo{person}{Jenya Lee}, \bibinfo{person}{Diana Liskovich}, \bibinfo{person}{Yinghai Lu}, \bibinfo{person}{Yuning Mao}, \bibinfo{person}{Xavier Martinet}, \bibinfo{person}{Todor Mihaylov}, \bibinfo{person}{Pushkar Mishra}, \bibinfo{person}{Igor Molybog}, \bibinfo{person}{Yixin Nie}, \bibinfo{person}{Andrew Poulton}, \bibinfo{person}{Jeremy Reizenstein}, \bibinfo{person}{Rashi Rungta}, \bibinfo{person}{Kalyan Saladi}, \bibinfo{person}{Alan Schelten}, \bibinfo{person}{Ruan Silva}, \bibinfo{person}{Eric~Michael Smith}, \bibinfo{person}{Ranjan Subramanian}, \bibinfo{person}{Xiaoqing~Ellen Tan}, \bibinfo{person}{Binh Tang}, \bibinfo{person}{Ross Taylor}, \bibinfo{person}{Adina Williams}, \bibinfo{person}{Jian~Xiang Kuan}, \bibinfo{person}{Puxin Xu}, \bibinfo{person}{Zheng Yan}, \bibinfo{person}{Iliyan Zarov}, \bibinfo{person}{Yuchen
  Zhang}, \bibinfo{person}{Angela Fan}, \bibinfo{person}{Melanie Kambadur}, \bibinfo{person}{Sharan Narang}, \bibinfo{person}{Aurelien Rodriguez}, \bibinfo{person}{Robert Stojnic}, \bibinfo{person}{Sergey Edunov}, {and} \bibinfo{person}{Thomas Scialom}.} \bibinfo{year}{2023}\natexlab{}.
\newblock \bibinfo{title}{Llama 2: Open Foundation and Fine-Tuned Chat Models}.
\newblock
\newblock
\showeprint[arxiv]{2307.09288}


\bibitem[Vaswani et~al\mbox{.}(2017)]%
        {LLM_1}
\bibfield{author}{\bibinfo{person}{Ashish Vaswani}, \bibinfo{person}{Noam Shazeer}, \bibinfo{person}{Niki Parmar}, \bibinfo{person}{Jakob Uszkoreit}, \bibinfo{person}{Llion Jones}, \bibinfo{person}{Aidan~N. Gomez}, \bibinfo{person}{\L{}ukasz Kaiser}, {and} \bibinfo{person}{Illia Polosukhin}.} \bibinfo{year}{2017}\natexlab{}.
\newblock \showarticletitle{Attention is all you need}. In \bibinfo{booktitle}{\emph{Proceedings of the 31st International Conference on Neural Information Processing Systems}} \emph{(\bibinfo{series}{NIPS'17})}. \bibinfo{pages}{6000–6010}.
\newblock


\bibitem[Wang et~al\mbox{.}(2021)]%
        {mobicom_expert_quant}
\bibfield{author}{\bibinfo{person}{Manni Wang}, \bibinfo{person}{Shaohua Ding}, \bibinfo{person}{Ting Cao}, \bibinfo{person}{Yunxin Liu}, {and} \bibinfo{person}{Fengyuan Xu}.} \bibinfo{year}{2021}\natexlab{}.
\newblock \showarticletitle{AsyMo: scalable and efficient deep-learning inference on asymmetric mobile CPUs}. In \bibinfo{booktitle}{\emph{Proceedings of the 27th Annual International Conference on Mobile Computing and Networking}} \emph{(\bibinfo{series}{MobiCom '21})}. \bibinfo{pages}{215–228}.
\newblock


\bibitem[Xia et~al\mbox{.}(2024)]%
        {quantllm}
\bibfield{author}{\bibinfo{person}{Haojun Xia}, \bibinfo{person}{Zhen Zheng}, \bibinfo{person}{Xiaoxia Wu}, \bibinfo{person}{Shiyang Chen}, \bibinfo{person}{Zhewei Yao}, \bibinfo{person}{Stephen Youn}, \bibinfo{person}{Arash Bakhtiari}, \bibinfo{person}{Michael Wyatt}, \bibinfo{person}{Donglin Zhuang}, \bibinfo{person}{Zhongzhu Zhou}, \bibinfo{person}{Olatunji Ruwase}, \bibinfo{person}{Yuxiong He}, {and} \bibinfo{person}{Shuaiwen~Leon Song}.} \bibinfo{year}{2024}\natexlab{}.
\newblock \showarticletitle{{Quant-LLM}: Accelerating the Serving of Large Language Models via {FP6-Centric} {Algorithm-System} {Co-Design} on Modern {GPUs}}. In \bibinfo{booktitle}{\emph{2024 USENIX Annual Technical Conference}} \emph{(\bibinfo{series}{USENIX ATC '24})}. \bibinfo{pages}{699--713}.
\newblock


\bibitem[Xiao et~al\mbox{.}(2023)]%
        {xiao2023smoothquant}
\bibfield{author}{\bibinfo{person}{Guangxuan Xiao}, \bibinfo{person}{Ji Lin}, \bibinfo{person}{Mickael Seznec}, \bibinfo{person}{Hao Wu}, \bibinfo{person}{Julien Demouth}, {and} \bibinfo{person}{Song Han}.} \bibinfo{year}{2023}\natexlab{}.
\newblock \showarticletitle{{S}mooth{Q}uant: Accurate and Efficient Post-Training Quantization for Large Language Models}. In \bibinfo{booktitle}{\emph{Proceedings of the 40th International Conference on Machine Learning}} \emph{(\bibinfo{series}{ICML '23})}.
\newblock


\bibitem[Yi et~al\mbox{.}(2023)]%
        {Yi2023EdgeMoEFO}
\bibfield{author}{\bibinfo{person}{Rongjie Yi}, \bibinfo{person}{Liwei Guo}, \bibinfo{person}{Shiyun Wei}, \bibinfo{person}{Ao Zhou}, \bibinfo{person}{Shangguang Wang}, {and} \bibinfo{person}{Mengwei Xu}.} \bibinfo{year}{2023}\natexlab{}.
\newblock \showarticletitle{EdgeMoE: Fast On-Device Inference of MoE-based Large Language Models}.
\newblock \bibinfo{journal}{\emph{ArXiv}}  \bibinfo{volume}{abs/2308.14352} (\bibinfo{year}{2023}).
\newblock


\bibitem[Zellers et~al\mbox{.}(2019)]%
        {hellaswag}
\bibfield{author}{\bibinfo{person}{Rowan Zellers}, \bibinfo{person}{Ari Holtzman}, \bibinfo{person}{Yonatan Bisk}, \bibinfo{person}{Ali Farhadi}, {and} \bibinfo{person}{Yejin Choi}.} \bibinfo{year}{2019}\natexlab{}.
\newblock \showarticletitle{{H}ella{S}wag: Can a Machine Really Finish Your Sentence?}. In \bibinfo{booktitle}{\emph{Proceedings of the 57th Annual Meeting of the Association for Computational Linguistics}} \emph{(\bibinfo{series}{ACL '19})}. \bibinfo{pages}{4791--4800}.
\newblock


\bibitem[Zhang et~al\mbox{.}(2022)]%
        {LLM_opt}
\bibfield{author}{\bibinfo{person}{Susan Zhang}, \bibinfo{person}{Stephen Roller}, \bibinfo{person}{Naman Goyal}, \bibinfo{person}{Mikel Artetxe}, \bibinfo{person}{Moya Chen}, \bibinfo{person}{Shuohui Chen}, \bibinfo{person}{Christopher Dewan}, \bibinfo{person}{Mona Diab}, \bibinfo{person}{Xian Li}, \bibinfo{person}{Xi~Victoria Lin}, \bibinfo{person}{Todor Mihaylov}, \bibinfo{person}{Myle Ott}, \bibinfo{person}{Sam Shleifer}, \bibinfo{person}{Kurt Shuster}, \bibinfo{person}{Daniel Simig}, \bibinfo{person}{Punit~Singh Koura}, \bibinfo{person}{Anjali Sridhar}, \bibinfo{person}{Tianlu Wang}, {and} \bibinfo{person}{Luke Zettlemoyer}.} \bibinfo{year}{2022}\natexlab{}.
\newblock \bibinfo{title}{OPT: Open Pre-trained Transformer Language Models}.
\newblock
\newblock
\showeprint[arxiv]{2205.01068}


\bibitem[Zhong et~al\mbox{.}(2024)]%
        {adapMoE}
\bibfield{author}{\bibinfo{person}{Shuzhang Zhong}, \bibinfo{person}{Ling Liang}, \bibinfo{person}{Yuan Wang}, \bibinfo{person}{Runsheng Wang}, {and} \bibinfo{person}{Meng~Li Ru~Huang}.} \bibinfo{year}{2024}\natexlab{}.
\newblock \showarticletitle{AdapMoE: Adaptive Sensitivity-based Expert Gating and Management for Efficient MoE Inference.}. In \bibinfo{booktitle}{\emph{IEEE International Conference on Computer-Aided Design}} \emph{(\bibinfo{series}{ICCAD '24})}.
\newblock


\bibitem[Zhu et~al\mbox{.}(2024)]%
        {llama-moe}
\bibfield{author}{\bibinfo{person}{Tong Zhu}, \bibinfo{person}{Xiaoye Qu}, \bibinfo{person}{Daize Dong}, \bibinfo{person}{Jiacheng Ruan}, \bibinfo{person}{Jingqi Tong}, \bibinfo{person}{Conghui He}, {and} \bibinfo{person}{Yu Cheng}.} \bibinfo{year}{2024}\natexlab{}.
\newblock \showarticletitle{LLaMA-MoE: Building Mixture-of-Experts from LLaMA with Continual Pre-training}.
\newblock \bibinfo{journal}{\emph{arXiv preprint arXiv:2406.16554}} (\bibinfo{year}{2024}).
\newblock


\end{thebibliography}

\end{document}